\documentclass[reqno]{amsart}

\usepackage{amsmath, amsthm, amssymb, epsfig, amscd}

\renewcommand{\d}{\delta}

\newcommand{\cstar}{\mathbb{C}^{\times}}
\newcommand{\Z}{\mathbb{Z}}
\newcommand{\C}{\mathbb{C}}
\newcommand{\R}{\mathbb{R}}
\newcommand{\Reals}{\mathbb{R}}
\newcommand{\cH}{\mathcal{H}}

\newcommand{\cS}{\mathcal{S}}

\newcommand{\cO}{\mathcal{O}}
\newcommand{\cL}{\mathcal{L}}
\newcommand{\E}{\mathcal{E}}
\newcommand{\K}{\mathcal{K}}
\newcommand{\U}{\mathcal{U}}
\newcommand{\tr}{\text{tr}}

\newcommand{\Fred}{\text{Fred}}

\newcommand{\GLtr}{GL_{\text{tr}}}
\newcommand{\Utr}{U_{\text{tr}}}
\newcommand{\ind}{\text{ind}}

\newcommand{\dimker}{\text{dimker}}

\newcommand{\End}{\text{End}}
\newcommand{\Mod}{\text{Mod}}
\newcommand{\ev}{\text{ev}}
\newcommand{\odd}{\text{odd}}

\theoremstyle{plain}
\newtheorem{proposition}{Proposition}[section]

\begin{document}

\title[Chern character in twisted K-theory]{Chern character in
twisted K-theory:\\ Equivariant and Holomorphic cases}

\author[V. Mathai]{Varghese Mathai}
\address[Varghese Mathai]
{Department of Pure Mathematics\\
University of Adelaide\\
Adelaide, SA 5005 \\
Australia}
\email{vmathai@maths.adelaide.edu.au }

\author[D. Stevenson]{Danny Stevenson}
\address[Danny Stevenson]
{Department of Pure Mathematics\\
University of Adelaide\\
Adelaide, SA 5005 \\
Australia}
\email{dstevens@maths.adelaide.edu.au}

\thanks{The authors acknowledge the support of the Australian
Research Council.}



\begin{abstract}
It was argued in \cite{Wit}, \cite{BM} that in the presence of a nontrivial
$B$-field, $D$-brane charges in type IIB string theories are
classified by
twisted $K$-theory. In \cite{bcmms}, it was proved that twisted $K$-theory
is canonically isomorphic to bundle gerbe $K$-theory, whose elements
are ordinary Hilbert bundles on a principal projective
unitary bundle, with an action of the bundle gerbe determined by
the principal projective
unitary bundle. The principal projective
unitary bundle is in turn determined by the twist.
This paper studies in detail the Chern-Weil representative of
the Chern character of bundle gerbe
$K$-theory that was introduced in
\cite{bcmms}, extending the construction to the equivariant and the
holomorphic cases. Included is a discussion of interesting
examples.
\end{abstract}
\maketitle

\section{Introduction}
\label{sec:one}

As argued by Minasian-Moore \cite{MinMoo}, Witten \cite{Wit},
$D$-brane charges in type IIB
string theories are classified by $K$-theory, which arises
from the fact that $D$-branes have
vector bundles on their world-volumes.
In the presence of a nontrivial $B$-field but whose
Dixmier-Douady class is a torsion element of $H^3(M, \Z)$, Witten also
showed that $D$-branes no longer have honest vector bundles on their
world-volumes, but they have a twisted or gauge bundle. These are
vector bundle-like objects whose transition functions
$g\sb {ab}$ on triple overlaps satisfy $$g_{ab}g_{bc}g_{ca}\:\,=\,\:
h_{abc}I,$$ where $h\sb {abc}$ is a Cech 2-cocycle representing
an element of $H^2(M, \underline{{ U}(1)})\cong H^3(M,\Z)$, and
$I$ is the $n\times n$ identity matrix, for $n$ coincident D-branes.
This proposal was later related by Kapustin \cite{Ka} to projective
modules over  Azumaya algebras.

In the presence of a nontrivial $B$-field whose
Dixmier-Douady class is a general element of $H^3(M, \Z)$, it was
proposed in \cite{BM}  that
$D$-brane charges in type IIB
string theories are measured by the twisted $K$-theory that was
described by Rosenberg \cite{Ros}, and the twisted bundles
on the $D$-brane world-volumes were elements in this twisted
$K$-theory. In \cite{bcmms}, it was shown
that these twisted bundles are equivalent to ordinary Hilbert bundles on
the total  space of the principal projective unitary bundle $P$ over $M$
with  Dixmier-Douady invariant $[h_{abc}]\in H^3(M,\Z)$. These
Hilbert bundles over $P$ are in addition required to have an action of
the bundle gerbe associated to $P$,
and were called bundle gerbe modules. That is, bundle
gerbe modules on $M$ are vector bundles over $P$ that are
invariant under a projective action of the projective unitary group.
The theory of bundle gerbes was initated by Murray \cite{Mur}
as a bundle theory analogue of the theory of gerbes due to Giraud
\cite{Gi} and Brylinski \cite{Bry} that used sheaves of categories.
It was also proved in
\cite{bcmms} that the $K$-theory defined using bundle gerbe modules, called
{\em bundle gerbe $K$-theory,
is naturally isomorphic to twisted $K$-theory} as defined
by Rosenberg \cite{Ros}, thus yielding a nice geometric
interpretation of twisted bundles on the $D$-brane world-volumes
in the presence of a nontrivial $B$-field.
There are discussions in \cite{DMW},
\cite{MalMooSei} on the uses of twisted $K$-theory in string
theory.

This paper extends the Chern-Weil construction of the Chern character
of bundle-gerbe $K$-theory that was defined in
\cite{bcmms}, to the equivariant and the holomorphic cases.
It also details the tensor product construction in
bundle gerbe $K$-theory, which turns out to be a delicate
matter when the curvature of the $B$-field is non-trivial.
The non-trivial multiplicativity property of the Chern character
is also studied, as well as the expression of the Chern character
in the odd degree case.  It has been noted by Witten \cite{Wit}
that $D$-brane charges in Type IIA string theories are classified
by $K^1(M)$ with appropriate compact support conditions.
The relevance of the equivariant case to
conformal field theory was highlighted by the remarkable
discovery by Freed, Hopkins and Teleman \cite{Fre} that
the twisted $G$-equivariant $K$-theory of a compact
connected Lie group
$G$ (with mild hypotheses) is graded isomorphic to the Verlinde
algebra of $G$, with a shift given by the dual Coxeter number
and the curvature of the $B$-field.
Recall that
Verlinde algebra of  a compact connected Lie group $G$ is defined in
terms of positive energy  representations of the loop group of $G$,
and arises naturally in  physics in Chern-Simons theory which is
defined using quantum groups  and conformal field theory.
The relevance  of aspects of holomorphic $K$-theory to physics
have been discussed in  Sharpe \cite{Sh} and Kapustin-Orlov
\cite{KaOr}, but using coherent sheaves and categories, instead
of holomorphic vector bundles that is used in this paper.

\S2 contains a brief review of the theory of bundle gerbes and its
equivariant analogue, as well as the theory of connections and
curvature on these. In \S3 a more detailed account of twisted
cohomology is given than what appears in  \cite{bcmms}. \S4
deals with bundle gerbe $K$-theory and
the delicate problem of defining the tensor product as well
as the multiplicativity property  of the Chern
character in this context. \S5 contains a derivation of the
Chern character in odd degree bundle gerbe $K$-theory. \S6
contains the extension of
the earlier discussion of the Chern character to the case of
equivariant bundle gerbe $K$-theory, and in \S7 to the case of
holomorphic bundle gerbe $K$-theory. \S8 contains a discussion
of the natural class of examples, of $Spin$ and
$Spin^{\mathbb C}$ bundle gerbes, and the associated
spinor bundle gerbe modules. Also included are examples in
the holomorphic and equivariant cases.
We would like to thank Eckhard Meinrenken for pointing 
out an error in an earlier version of this paper.  We would 
also like to thank the referee for useful comments.

\section{Review of bundle gerbes}
\label{sec:two}

Bundle gerbes were introduced by Murray \cite{Mur}
and provided an alternative to Brylinski's category
theoretic notion of a gerbe \cite{Bry}.  Gerbes and bundle
gerbes provide a geometrical realisation of
elements of $H^3(M,\Z)$ just as line bundles
provide a geometrical realisation of $H^2(M,\Z)$.
We briefly review here the definition and
main properties of bundle gerbes.

A bundle gerbe on a manifold $M$ consists of a
submersion $\pi\colon Y\to M$ 
together with a line bundle $L\to Y^{[2]}$
on the fibre product $Y^{[2]} = Y\times_{\pi}Y$.
$L$ is required to have a product, that is a line
bundle isomorphism which on the fibres takes the
form
\begin{equation}
\label{eq:bundle gerbe product}
L_{(y_1,y_2)}\otimes L_{(y_2,y_3)}\to L_{(y_1,y_3)}
\end{equation}
for points $y_1$, $y_2$ and $y_3$ all belonging to the
same fibre of $Y$.  This product is required to be
associative in the obvious sense.  

A motivating example of a bundle gerbe arises
whenever we have a central extension of groups
$\cstar\to \hat{G}\to G$ and a principal $G$ bundle
$P\to M$ on $M$.  Then on the fibre product
$P^{[2]}$ we can form the map $g\colon P^{[2]}\to G$
by comparing two points which lie in the same fibre.
We can use this map $g$ to pull back the line bundle
associated to the principal $\cstar$ bundle $\hat{G}$.
The resulting line bundle $L\to P^{[2]}$ (the
`primitive' line bundle) is a bundle gerbe with the
product induced by the product in the group $\hat{G}$.
In \cite{Mur} this bundle gerbe is called  
the lifting bundle gerbe.
We will be particularly interested in the case
when $G = PU$, the projective unitary group of some
separable Hilbert space $\cH$.  Then, as is well
known, we have the central extension $U(1)\to U\to PU$.
Therefore, associated to any principal $PU$ bundle
on $M$ we have an associated bundle gerbe.

For use in Section~\ref{sec:seven} we will explain here
what we mean by a $G$-equivariant bundle gerbe.
Suppose that $G$ is a compact Lie group acting
on the manifold $M$.  We first recall the
notion of a $G$-equivariant line bundle on $M$
(see for instance \cite{Bry}).  We denote by
$p_1$ and $m$ the maps $M\times G\to M$ given
by projection onto the first factor and the action
of $G$ on $M$ respectively.  A line bundle
$L\to M$ is said to be $G$-equivariant if there
is a line bundle isomorphism $\sigma\colon p_1^*L\to m^* L$
covering the identity on $M\times G$. Thus
$\sigma$ is equivalent to the data of a family of
maps which fiberwise are of the form
\begin{equation}
\label{eq:G-equivariant line bundle}
\sigma_g\colon L_m \to L_{g(m)}
\end{equation}
for $m\in M$ and $g\in G$.  These maps are required
to vary smoothly with $m$ and $g$ and satisfy the obvious
associativity condition.
A $G$-equivariant bundle
gerbe consists of the data of a  
submersion $\pi\colon Y\to M$ such that $Y$ has a
$G$-action covering the action on $M$ together with
a $G$-equivariant line bundle $L\to Y^{[2]}$.
$L\to Y^{[2]}$
has a product, i.e. a $G$-equivariant line bundle
isomorphism taking the form~(\ref{eq:bundle gerbe product})
on the fibres.  We remark that this definition is really
too strong, we should require that $G$ acts on $L$ only
up to a `coherent natural transformation'.  However the
definition we have given will be
more than adequate for our purposes.

A bundle gerbe connection $\nabla$ on a bundle gerbe
$L\to Y^{[2]}$ is a connection on the line bundle
$L$ which is compatible with the product~(\ref{eq:bundle
gerbe product}), i.e. $\nabla(st) = \nabla(s)t +
s\nabla(t)$ for sections $s$ and $t$ of $L$.
In \cite{Mur} it is shown that bundle gerbe
connections always exist.  Let  
$F_{\nabla}$ denote the curvature of a bundle gerbe connection $\nabla$.    
It is shown in \cite{Mur} that we can find a
$2$-form $f$ on $Y$ such that $F_{\nabla} = \delta(f) =
\pi_2^*f - \pi_1^*f$.  $f$ is unique up to $2$-forms
pulled back from $M$.  A choice of $f$ is called a
choice of a curving for $\nabla$.  Since $F_{\nabla}$ is
closed we must have $df = \pi^*\omega$ for some necessarily
closed $3$-form $\omega$ on $M$.  $\omega$ is called
the $3$-curvature of the bundle gerbe connection $\nabla$
and curving $f$.  It is shown in \cite{Mur} that
$\omega$ has integral periods.

For later use it will also be of interest to know that
we can find a $G$-equivariant connection on a $G$-equivariant
bundle gerbe $L\to Y^{[2]}$.  This is a connection on
$L$ which is compatible with the product structure on
$L$ and is also invariant under the action of $G$.
Since $G$ is assumed to be compact, we can always find
such a connection by an averaging procedure.

\section{Twisted Cohomology}
\label{sec:three}

Twisted cohomology turns out to be the natural target
space for the Chern character defined in bundle gerbe
$K$-theory. In this section we give a short account of 
the main properties of twisted cohomology.  
Suppose $H$ is a closed differential $3$-form.
We can use $H$ to construct a differential $\delta_H$ on the
algebra $\Omega^\bullet(M)$ of differential forms
on $M$ by setting $\delta_H(\omega) = d\omega -H\omega$
for $\omega \in \Omega^\bullet(M)$.  It is easy to check
that indeed $\delta_H^2 = 0$.  
We set $H^\bullet(M,H)$ to be the quotient $\text{ker}\delta_H/
\text{im}\delta_H$.  If $\lambda$ is a differential
$2$-form on $M$ then we can form the differential
$\delta_{H+d\lambda}$ and the group $H^\bullet(M,H+d\lambda)$.
One can construct an isomorphism $H^\bullet(M,H) \to
H^\bullet(M,H+d\lambda)$ by sending a class in
$H^\bullet(M,H)$ represented by $\omega$ to the class in
$H^\bullet(M,H+d\lambda)$ represented by $\exp(\lambda)\omega$.
So any two closed differential $3$-forms $H$ and $H'$
representing the same class in $H^3(M,\Reals)$
determine isomorphic  
cohomology groups $H^\bullet(M,H)$ and $H^\bullet(M,H')$.
The two groups $H^\bullet(M,H)$ and
$H^\bullet(M,H')$ are not uniquely isomorphic, as
there are many $2$-forms $\lambda$ such that $H' = H + d\lambda$.
Note also that there is a homomorphism of groups
$H^\bullet(M,H)\otimes H^\bullet(M,H') \to
H^\bullet(M,H+H')$ defined by sending a class
$[\omega]\otimes [\rho]$ in $H^\bullet(M,H)\otimes
H^\bullet(M,H')$ to the class $[\omega\rho]$ in
$H^\bullet(M,H+H')$.  
Finally note that $H^\bullet(M,H)$ has a natural
structure of an $H^\bullet(M)$ module: if $[\omega] \in
H^\bullet(M,H)$ and $[\rho] \in H^\bullet(M)$ then
$[\rho\omega]\in H^\bullet(M,H)$.  If $P$ is a principal 
$PU$ bundle on $M$ with Dixmier-Douady class $\delta(P) 
\in H^3(M;\Z)$ then we define a group $H^\bullet(M;P)$ 
by choosing a closed $3$-form $H$ on $M$ representing 
the image of $\delta(P)$ in de Rham cohomology and set 
$H^\bullet(M;P) = H^\bullet(M;H)$.

\begin{proposition}
\label{prop:twisted cohomology}
Suppose that $P$ is a principal $PU(\cH)$ bundle on
$M$.  Then $H^\bullet(M,P)$ is an abelian group
with the following properties:
\begin{enumerate}

\item If the principal $PU(\cH)$ bundle $P$ is
trivial then there is an isomorphism $H^\bullet(M,P)
= H^\bullet(M)$.

\item If the principal $PU(\cH)$ bundle $P$ is isomorphic
to another principal $PU(\cH)$ bundle $Q$ then there
is an isomorphism of groups $H^\bullet(M,P) =
H^\bullet(M,Q)$.

\item If $Q$ is another principal $PU(\cH)$ bundle on $M$,
then there is a homomorphism $H^\bullet(M,P)\otimes
H^\bullet(M,Q) \to H^\bullet(M,P\otimes Q)$.

\item $H^\bullet(M,P)$ has a natural structure as a $H^\bullet(M)$
module.

\item If $f\colon N\to M$ is a smooth map then there is
an induced homomorphism $f^*\colon H^\bullet(M,P) \to
H^\bullet(N,f^{*}P)$.
\end{enumerate}
\end{proposition}

In $3$ above the $PU$ bundle $P\otimes Q$ on $M$ obtained 
from $PU$ bundles $P$ and $Q$ is defined \cite{Bry} by 
first forming the $PU\times PU$ bundle $P\times_M Q$ 
on $M$ and then using the homomorphism $PU(\cH)\times 
PU(\cH)\to PU(\cH\otimes\cH)$ to define $P\otimes Q 
= (P\times_M Q)\times_{PU\times PU}PU$.  We can then choose 
an isometry $\cH\otimes\cH = \cH$ to make $P\otimes Q$ 
into a principal $PU(\cH)$ bundle.

\section{The Twisted Chern Character: Even Case}
\label{sec:four}

\subsection{Geometric models of $\tilde{K}^0(X)$} 

We first discuss some features of reduced even $K$-theory 
$\tilde{K}^0(X)$.    
Since $\tilde{K}^0(X)$ is a ring, it follows that a model
for the classifying space of $\tilde{K}^0(X)$ must be a
`ring space' in the sense that there exist two $H$-space
structures on the model classifying space which satisfy
the appropriate distributivity axioms.  
Let $\Fred_0$ denote
the connected component of the identity in the
space of Fredholms $\Fred$, and  $BGL_{\K}$ 
the classifying space of the group of
invertible operators $GL_{\K}$ which differ from the identity
by a compact operator.
The $H$-space
structures on $\Fred_0$ and on $BGL_{\K}$ which induce
addition in $\tilde{K}^0(X)$ are easy to describe; on
$\Fred_0$ the $H$-space structure is given by composition
of Fredholm operators while on $BGL_{\K}$ the $H$-space
structure is given by the group multiplication on
$BGL_{\K}$.  The $H$-space structures inducing multiplication
on $\tilde{K}^0(X)$ are harder to describe.
On $\Fred_0$ this $H$-space structure is given as follows
(see \cite{AtiSin}).  If $F_1$ and $F_2$
are Fredholm operators on the separable Hilbert space
$\cH$ then we can form the tensor product operator
$F_1\otimes I + I\otimes F_2$.  One can show that this
operator is Fredholm of index $\dimker(F_1)\dimker(F_2)
- \dimker(F_1^*)\dimker(F_2^*)$.  In particular, if
$\ind(F_1) = \ind(F_2) = 0$ then $F_1\otimes I +
I\otimes F_2$ is Fredholm of index zero on $\cH\otimes
\cH$.  If we choose an isometry of $\cH\otimes \cH$ with
$\cH$ then this operation of tensor product of Fredholms
induces an $H$-map on $\Fred_0$ which one can show
corresponds to the multiplication on $\tilde{K}^0(X)$.
Note that this works only for $\Fred_0$, for $\Fred$
one must use a $\Z_2$-graded version of this tensor
product to get the right $H$-map.  Since $BGL_{\K}$ and
$\Fred_0$ are homotopy equivalent there is an induced
$H$-map on $BGL_{\K}$, however it is difficult to see
what this map is at the level of principal $GL_{\K}$
bundles.

To investigate this $H$-map on $BGL_{\K}$, we shall
replace principal $GL_{\K}$ bundles $P\to X$ with
Hilbert vector bundles $E\to X$ equipped with a fixed reduction
of the structure group of $E$ to $GL_{\K}$.  We shall
refer to such vector bundles as $GL_{\K}$-vector bundles.
Note that Koschorke, in \cite{Kos}, reserves the
terminology $GL_{\K}$-vector bundle
for more general objects.  
In other
words we can find local trivialisations $\phi_U\colon
E|_U \to U\times \cH$ such that the transition functions
$g_{UV}$ relative to these local trivialisations take
values in $GL_{\K}$.  Another way of looking at this is
that the principal frame bundle $F(E)$ of $E$ has a reduction of
its structure group from $GL$ to $GL_{\K}$ (note that
there will be many such reductions).  This reduction is determined
up to isomorphism by a classifying map $X\to \Fred_0$.
Suppose we are given two $GL_{\K}$-vector bundles $E_1$ and $E_2$
on $X$ with the
reductions of the frame bundles $F(E_1)$ and $F(E_2)$
to $GL_{\K}$ corresponding to maps $F_1,F_2\colon X\to
\Fred_0$.  
We want to know the relation of the pullback of the 
universal $GL_{\K}$ bundle over $\Fred_0$ to the bundles 
$F(E_1)$ and $F(E_2)$.  We can suppose that $X$ is covered 
by open sets $U$ small enough so that we can write 
$F(x) = G_U(x) + K_U(x)$, $F_1(x) = g_U(x) + k_U(x)$ and 
$F_2(x) = g'_U(x) + k'_U(x)$ where $G_U$, $g_U$ and 
$g'_U$ are invertible and $K_U$, $k_U$ and $k'_U$ are 
compact --- see for example \cite{Ati}.  
It is then not hard to show that we can find 
$h_U\colon U\to GL$ so that $G_{UV} = h_U^{-1}g_{UV}\otimes 
g'_{UV}h_V$ on $U\cap V$ where $G_{UV} = G_U^{-1}G_V$, 
$g_{UV} = g_Ug_V^{-1}$ and $g'_{UV} = g^{\prime -1}_Ug'_V$.  
The $g_{UV}$ and $g'_{UV}$ are the $GL_{\K}$ 
valued transition functions for the $GL_{\K}$ bundles 
$F(E_1)$ and $F(E_2)$ while the $G_{UV}$ are the $GL_{\K}$ 
valued transition functions for the pullback by $F$ of the 
universal $GL_{\K}$ bundle on $\Fred_0$.  It follows that 
this last bundle is a reduction of the structure group to $GL_{\K}$ of the 
frame bundle $F(E_1\otimes E_2)$ of the tensor product 
Hilbert bundle $E_1\otimes E_2$.    
We therefore have an alternative description of the
ring $\tilde{K}^0(X)$
as the Grothendieck group associated to the semi-group
$V_{GL_{\K}}(X)$ of $GL_{\K}$-vector bundles on $X$, where
the addition and multiplication in $V_{GL_{\K}}(X)$ are
given by direct sum and tensor product of $GL_{\K}$-vector
bundles, after identifying $\cH\oplus\cH$, $\cH\otimes
\cH$ and $\cH$ via isometries.

\subsection{Twisted $K$-theory and bundle gerbe modules} 

After these preliminary remarks, we now turn to 
the definition of twisted $K$-theory.  
Given a principal $PU$ bundle $P$ on $M$ with 
Dixmier-Douady class $\delta(P)\in H^3(M;\Z)$,  
Rosenberg \cite{Ros} defines twisted $K$-groups
$K^i(M,P)$ to be the $K$-groups $K_i(A(P))$ of the
algebra $A(P) = C^{\infty} (\E,M)$.  Here $\E = P\times_{PU}\K$ 
is the bundle on $M$ associated to $P$ via the adjoint action 
of $PU$ on the compact operators $\K$.  
Rosenberg shows that one can identify 
$K^0(M,P)$ with the set of homotopy classes of
$PU$-equivariant maps $[P,\Fred]^{PU}$ and
$K^1(M,P)$ with the set of homotopy classes of
$PU$-equivariant maps $[P,GL_{\K}]^{PU}$.  We 
do not assume that the class $\delta(P)$ is necessarily 
torsion.  The reduced theory
$\tilde{K}^0(M,P)$ can be shown to be equal to
$[P,\Fred_0]^{PU}$.  We can replace $\Fred_0$
by any other homotopy equivalent space on which
$PU$ acts.  For example we could take $BGL_{\K}$ instead
of  $\Fred_0$. Various other characterisations
of $\tilde{K}^0(M,P)$ are possible, we summarise
them in the following proposition from \cite{bcmms}

\begin{proposition}[\cite{bcmms}]
\label{prop:various twisted K-theory definitions}
Given a principal $PU$ bundle $P\to M$ with Dixmier-Douady
class $\delta(P)\in H^3(M,\Z)$ we have the following
isomorphisms of groups:
\begin{enumerate}
\item $\tilde{K}^0(M,P)$
\item space of homotopy classes of sections of
the associated bundle $P\times_{PU} BGL_{\K}$
\item space of homotopy classes of $PU$
equivariant maps: $[P,BGL_{\K}]^{PU}$.
\item space of isomorphism classes of $PU$
covariant $GL_{\K}$ bundles on $P$.
\item space of homotopy classes of Fredholm
bundle maps $F\colon \cH_0 \to \cH_1$ between
Hilbert vector bundles $\cH_0$ and $\cH_1$
on $P$ which are bundle gerbe modules for the lifting bundle gerbe
$L\to P^{[2]}$.
\end{enumerate}
\end{proposition}
Here a $PU$ covariant $GL_{\K}$ bundle $Q$ on $P$
is a principal $GL_{\K}$ bundle $Q\to P$ together
with a right action of $PU$ on $Q$ covering the
action on $P$ such that $(qg)[u] = q[u]u^{-1}gu$
where $g\in GL_{\K}$, $[u] \in PU$. 
In 5 above (see below for the definition of a 
bundle gerbe module) we require that the map $F\colon
\cH_0 \to \cH_1$ is a Fredholm map on the fibres
and is also compatible with the module structures
of $\cH_0$ and $\cH_1$ in the sense that the
following diagram commutes:
$$
\begin{array}{ccc}
\pi_1^*\cH_0 \otimes L &
\stackrel{\pi_1^*F\otimes 1}{\rightarrow}
& \pi_1^*\cH_1 \otimes L                    \\
\downarrow & & \downarrow                    \\
\pi_2^*\cH_0 & \stackrel{\pi_2^*F}{\rightarrow}
& \pi_2^*\cH_1                                \\
\end{array}
$$
Given two such maps $F,G\colon \cH_0 \to \cH_1$
we require that they are homotopic via a homotopy
which preserves the module structures of $\cH_0$
and $\cH_1$ in the above sense.
It can be
shown that twisted $K$-theory has the following
properties.
\begin{proposition}
\label{prop:twisted K-theory properties}
Twisted $K$-theory satisfies the following
properties:
\begin{enumerate}
\item If the principal $PU$ bundle $P\to M$  
is trivial then $K^p(M,P) = K^p(M)$.
\item $K^p(M,P)$ is a module over $K^0(M)$.
\item If $P\to M$ and $Q\to M$ are principal 
$PU$ bundles on $M$ then there is a homomorphism  
$$
K^p(M,P)\otimes K^q(M,Q)\to K^{p+q}(M,P\otimes Q).    
$$
\item If $f\colon
N\to M$ is a map then there is a homomorphism
$$
K^p(M,P)\to K^p(M,f^*P),
$$
where $f^*P\to M$ denotes the pullback principal 
$PU$ bundle.  
\end{enumerate}
The reduced theory $\tilde{K}^\bullet(M,P)$
satisfies the analogous properties.
\end{proposition}

Associated to the principal $PU$ bundle $P\to M$ via
the central extension of groups $U(1)\to U\to PU$ is
the lifting bundle gerbe $L\to P^{[2]}$.  In \cite{bcmms}
the notion of a bundle gerbe module for $L$ was introduced.
When the Dixmier-Douady class $\delta(P)$ is not necessarily torsion
a bundle gerbe module for $L$ was defined to mean
a $GL_{\K}$-vector bundle $E$ on $P$ together with
an action of $L$ on $E$.  This was a vector bundle isomorphism
$\pi_1^*E\otimes L\to \pi_2^*E$ on $P^{[2]}$
which was compatible with the product on $L$.  We also
require a further condition relating to the action of
$U$ on the principal $GL_{\K}$ bundle associated to $E$.
More specifically, let $GL(E)$ denote the principal
$GL(\cH)$ bundle on $P$ associated to $E$ whose fibre at
a point $p$ of $P$ consists of all isomorphisms $f\colon \cH\to E_p$.
$g\in GL(\cH)$ acts via $fg = f\circ g$.  Since
$E$ has structure group $GL_{\K}$ there is a reduction
of $GL(E)$ to a principal $GL_{\K}$ bundle $R\subset GL(E)$.
We require that if $u\in U(\cH)$ and $p_2 = p_1[u]$, so that
$u\in L_{(p_1,p_2)}$, then the map $GL(E)_{p_1}\to GL(E)_{p_2}$
which sends $f\in GL(E)_{p_1}$ to $ufu^{-1}$ preserves $R$.
So if $f\in R_{p_1}$ then $ufu^{-1} \in R_{p_2}$.
A bundle gerbe module $E$ is not quite a $PU$-equivariant
vector bundle, since the action of $PU$ will not
preserve the linear structure on the fibres of $E$, note however
that the projectivisation of $E$, $P(E)$, will descend to
a bundle of projective Hilbert spaces on $M$.
Recall from Section \ref{sec:four} that
the space of $GL_{\K}$ bundles on $P$ forms a semi-ring
under the operations of direct sum and tensor product.
It is easy to see that the operation of direct sum is compatible
with the action of the lifting bundle gerbe $L\to P^{[2]}$
and so the set of bundle gerbe modules for $L$,
$\Mod_{GL_{\K}}(L, M)$, has a natural structure as a semi-group.
We denote the group associated to $\Mod_{GL_{\K}}(L, M)$
by the Grothendieck construction by $\Mod_{GL_{\K}}(L, M)$ as well.
We remark as above that we can replace $BGL_{\K}$
by any homotopy equivalent space; in particular we
can consider $\GLtr$-vector bundles in place of
$GL_{\K}$-vector bundles.  To define characteristic
classes we must make this replacement.
The following result is proven in \cite{bcmms}.

\begin{proposition}[\cite{bcmms}]
\label{prop:bg K-theory = twisted K-theory}
If $L\to P^{[2]}$ is the lifting bundle gerbe for a
principal $PU$ bundle $P\to M$ with Dixmier-Douady class
$\delta(P)\in H^3(M,\Z)$ then
$$
\tilde{K}^0(M,P) = \Mod_{GL_{\K}}(L, M).
$$
The result remains valid when we replace $GL_{\K}$-vector
bundles by $GL_{\mathrm{tr}}$-vector bundles.
\end{proposition}

\subsection{The twisted Chern character in the even case} 

In \cite{bcmms} a homomorphism $ch_{P}\colon \tilde{K}^0
(M,P)\to H^{\ev}(M,P)$ was constructed with the
properties that 1) $ch_{P}$ is natural with respect
to pullbacks, 2) $ch_{P}$ respects the $\tilde{K}^0(M)$-module
structure of $\tilde{K}^0(M,P)$ and 3) $ch_{P}$ reduces
to the ordinary Chern character in the untwisted case
when $\delta(P) = 0$.  It was proposed that $ch_{P}$ was the
Chern character for (reduced) twisted $K$-theory.  We
review the construction of $ch_{P}$ here and prove
the result stated in \cite{bcmms} that $ch_{P}$ respects
the $\tilde{K}^0(M)$-module structure of $\tilde{K}^0(M,P)$.

To motivate the construction of $ch_{P}$, let
$P\to M$ be a principal $PU$ bundle with
Dixmier-Douady class $\delta(P)\in H^3(M,\Z)$ and
let $E\to P$ be a module for the lifting
bundle gerbe $L\to P^{[2]}$.  We suppose that
$L$ comes equipped with a bundle gerbe connection
$\nabla_L$ and a choice of curving $f$ such that
the associated $3$-curvature is $H$, a closed,
integral $3$-form on $M$ representing the image,
in real cohomology, of the Dixmier-Douady class
$\delta(P)$ of $P$.  Recall that
$L$ acts on $E$ via an isomorphism $\psi\colon
\pi_1^*E\otimes L\to \pi_2^*E$.  Since
the ordinary Chern character $ch$ is multiplicative,
we have
\begin{equation}
\label{eq:character equation}
\pi_1^*ch(E)ch(L) = \pi_2^*ch(E).
\end{equation}
Assume for the moment that this equation holds
on the level of forms.  Then $ch(L)$ is represented
by the curvature $2$-form $F_L$ of the
bundle gerbe connection $\nabla_L$ on $L$.  A choice
of a curving for $\nabla_L$ is a $2$-form $f$ on $P$
such that $F_L = \delta(f) = \pi_2^*f -
\pi_1^*f$.  It follows that $ch(L)$ is represented
by $\exp(F_L) = \exp(\pi_2^*f - \pi_1^*f) =
\exp(\pi_2^*f)\exp(-\pi_1^*f)$.  Therefore we can rearrange
the equation~(\ref{eq:character equation}) above to
get
$$
\pi_1^*\exp(-f)ch(E) = \pi_2^*\exp(-f)ch(E).
$$
Since we are assuming that this equation holds at the
level of forms, this implies that the form
$\exp(-f)ch(E)$ descends to to a form on $M$ which
is clearly closed with respect to the twisted
differential $d-H$.  To make this argument rigorous,
we need to choose connections on the module $E$ so
that the equation~(\ref{eq:character equation}) holds
on the level of forms.  To do this we need the notion of
a bundle gerbe module connection.
A bundle gerbe module connection on $E$ is a connection
$\nabla_E$ on the vector bundle $E$ which is compatible
with the bundle gerbe connection $\nabla_L$ on $L$ under the
action of $L$ on $E$.  In other words, under the
isomorphism $\psi\colon \pi_1^*E\otimes L\to \pi_2^*E$, we
have the transformation law
\begin{equation}
\label{eq:module connection}
\pi_1^*\nabla_E \otimes I + I\otimes \nabla_L =
\psi^{-1}\pi_2^*\nabla_E \psi.
\end{equation}
Let $F_L$ denote the curvature of $\nabla_L$
and let $F_E$ denote the curvature of $\nabla_E$.
Then the equation~(\ref{eq:module connection}) implies
that the curvatures satisfy the following equation:
$$
\pi_1^*F_E + F_L = \psi^{-1}\pi_2^*F_E \psi.
$$
Recall that the curving $f$ for the bundle gerbe
connection $\nabla_L$ satisfies $F_L = \d(f) =
\pi_2^*f - \pi_1^*f$.  Therefore we can rewrite the
equation above as
\begin{equation}
\label{eq:curvature transfm law}
\pi_1^*(F_E + fI) = \psi^{-1}\pi_2^* (F_E + fI)\psi.
\end{equation}
We would like to be able to take traces of powers of $F_E +fI$.
Then equation~(\ref{eq:curvature transfm law}) would imply
that the forms $\tr(F_E + fI)^p$ descend to $M$.
Unfortunately it is not possible to find connections
$\nabla_E$ so that the bundle valued $2$-form $F_E +fI$
takes values in the sub-bundle of trace class endomorphisms
of $E$ unless the $3$-curvature $H$ of the bundle
gerbe connection $\nabla_L$ and curving $f$ is zero.
Since we are interested in $K$-theory and hence in
$\Z_2$-graded vector bundles we can get around this
difficulty by considering connections $\nabla$ on the
$\Z_2$-graded module $E = E_0\oplus E_1$ which are of
the form $\nabla = \nabla_0 \oplus \nabla_1$, where
$\nabla_0$ and $\nabla_1$ are module connections on
$E_0$ and $E_1$ respectively such that the difference
$\nabla_0 - \nabla_1$ takes trace class values.
By this we mean that we can cover $P$ by open sets over
which $E_0$ and $E_1$ are trivial and in these local trivialisations
the connections $\nabla_0$ and $\nabla_1$ are given by
$d + A_0$ and $d+A_1$ respectively such that the difference
$A_0 - A_1$ is trace class. 

To see that we can always find such connections recall, 
as pointed out in \cite{bcmms}, that a $PU$-covariant 
$\GLtr$-bundle $Q$ on $P$ can be viewed as the total 
space $Q$ of a principal $GL_{\tr}\rtimes PU$ bundle 
over $M$ where the semi-direct product is defined by 
$(g_1,u_1)\cdot (g_2,u_2) = (\hat{u}_2^{-1}g_1\hat{u}_2g_2,u_1u_2)$.  
It follows that the Lie bracket on $Lie(GL_{\tr}\rtimes PU)$ is given by 
$$
[(\xi,U),(\eta,V)] = ([\xi,\eta] + [\xi,\hat{V}] + 
[\hat{U},\eta], [U,V]) 
$$
where $\xi$, $\eta\in Lie(GL_{\tr})$, $U,V\in Lie(PU)$ 
and $\hat{U}$, $\hat{V}$ are lifts of $U$ and $V$ to 
$Lie(U)$.  Let $\Theta$ be a connection $1$-form on the 
$GL_{\tr}\rtimes PU$ bundle $Q\to M$.  Let $p_1$ and 
$p_2$ denote the projections of $Lie(GL_{\tr}\rtimes PU)$ 
onto $Lie(GL_{\tr})$ and $Lie(PU)$ respectively.  We 
cannot define a connection $1$-form on the $GL_{\tr}$ bundle 
$Q\to P$ by $p_1(\Theta)$ as this $1$-form is not equivariant 
with respect to the action of $GL_{\tr}$.  Instead we have 
\begin{equation} 
\label{eq:transfm law for GLtr rtimes PU conn} 
p_1((g^{-1},1)\Theta(g,1)) = g^{-1}(p_1\Theta)g + 
g^{-1}\widehat{(p_2\Theta)}g - \widehat{(p_2\Theta)}. 
\end{equation} 
Note that $p_2\Theta$ pushes forward to define a 
connection $1$-form $A$ on the $PU$-bundle $P$.  
The transformation law~(\ref{eq:transfm law for GLtr rtimes PU conn}) 
turns out to be just what is needed to define a module 
connection on the associated bundle $E = Q\times_{GL_{\tr}}\cH$.  
Indeed, if $[s,f]$ is a section of $E$ 
then~(\ref{eq:transfm law for GLtr rtimes PU conn}) shows that 
\begin{equation} 
\label{eq:constructed module connection} 
\nabla_X[s,f] = [s,df(X) + s^*(p_1\Theta)(X)f + 
\sigma(A(X))f] 
\end{equation} 
is a well defined connection on $E$ (here $X$ is a vector 
field on $P$ and $\sigma\colon Lie(PU)\to Lie(U)$ is a 
choice of splitting of the central extension of Lie 
algebras $i\Reals \to Lie(U)\to Lie(PU)$).  To show that 
$\nabla$ defines a module connection on $E$ we want to show 
that under the isomorphism $\psi\colon \pi_1^{*}E\otimes 
L\to \pi_2^{*}E$ the connection $\pi_2^{*}\nabla$ on 
$\pi_2^{*}E$ is mapped into the tensor product connection 
$\pi_1^{*}\nabla +\nabla_L$ on $\pi_1^{*}E\otimes L$.  
$\psi$ sends a section $[s,f]\otimes [\hat{u},\lambda]$ 
of $\pi_1^{*}E\otimes L$ to the section $[su^{-1},\lambda 
\hat{u}f]$ of $\pi_2^{*}E$.   
Then we have 
\begin{eqnarray*} 
u\cdot \pi_2^{*}\nabla[su^{-1},\lambda \hat{u}f] & = & 
\pi_1^{*}\nabla[s,f]\otimes [\hat{u},\lambda] + [s,f] 
\otimes [\hat{u},d\lambda + \lambda(\hat{u}^{-1}d\hat{u} \\ 
&  & -\sigma(u^{-1}du)) + \lambda(\hat{u}^{-1}\sigma(\pi_2^*A)\hat{u} - 
\sigma(u^{-1}\pi_2^*Au)] 
\end{eqnarray*} 
One can check that if we define a connection $\nabla_L$ on 
the lifting bundle gerbe $L\to P^{[2]}$ by 
$\nabla_L[\hat{u},\lambda] =[\hat{u},d\lambda + \lambda(\hat{u}^{-1} 
d\hat{u} - \sigma(u^{-1}du)) + \lambda(\hat{u}^{-1}\sigma(\pi_2^*A) 
\hat{u} - \sigma(u^{-1}\pi_2^*Au))]$ then $\nabla_L$ is a 
bundle gerbe connection.  Therefore $\nabla$ is a module connection 
on $E$.  Suppose now that $E_1$ and $E_2$ are $GL_{\tr}$ modules 
for $L$.  Then by definition the frame bundles $U(E_1)$ 
and $U(E_2)$ have $GL_{\tr}$ reductions $Q_1$ and 
$Q_2$ respectively which are $PU$ covariant.  Therefore we 
may construct module connections on $E_1$ and $E_2$ by 
the above recipe.  We may choose connection 
$1$-forms $\Theta_1$ and $\Theta_2$  
on the $GL_{\tr}\rtimes PU$ bundles $Q_1$ and 
$Q_2$ respectively such that $p_2\Theta_1$ and 
$p_2\Theta_2$ push forward to define the same 
connection $1$-form $A$ on the $PU$ bundle $P$.  
It follows therefore from~(\ref{eq:constructed module connection}) 
that the difference of the associated module connections 
$\nabla_1$ and $\nabla_2$ in local trivialisations 
is trace class.  

We let $F_0$ and $F_1$ denote the curvatures of the
module connections $\nabla_0$ and $\nabla_1$
respectively.  It follows that the difference
$(F_0 + fI) - (F_1 + fI)$ and hence $(F_0 + fI)^p -
(F_1 + fI)^p$ takes trace class values.  It is shown
in \cite{bcmms} that the $2p$-forms $\tr((F_0 + fI)^p
- (F_1 + fI)^p)$ are defined globally on $P$ and moreover
they descend to $2p$-forms on $M$.  We have  
\begin{equation}
\tr(\exp(F_0 + fI) - \exp(F_1 + fI)) = \exp(f)
\tr(\exp(F_0) - \exp(F_1)),
\end{equation}
where we note that there is a cancellation of the 
degree zero terms --- this is due to the fact that we 
are considering reduced twisted $K$-theory.  
We summarise this discussion in the following
proposition from \cite{bcmms}.
\begin{proposition}
Suppose that $E= E_0\oplus E_1$ is a $\Z_2$-graded
module for the lifting bundle gerbe $L\to P^{[2]}$,
representing a class in $\tilde{K}^0(M,P)$ under the
isomorphism $\tilde{K}^0(M,P) = \Mod_{GL_{\mathrm{tr}}}(L, M)$.
Suppose that $\nabla = \nabla_0\oplus \nabla_1$
is a connection on the $\Z_2$-graded module $E$ such
that $\nabla_0$ and $\nabla_1$ are module connections
for $E_0$ and $E_1$ respectively such that the
difference $\nabla_0 - \nabla_1$ is trace class.
Let $ch_P(\nabla,E)$ denote the differential form on
$M$ whose lift to $P$ is equal to $\exp(f)\mathrm{tr}(\exp(F_0) -
\exp(F_1))$.  Then $ch_P(\nabla,E)$ is closed with
respect to the twisted differential $d-H$ on
$\Omega^\bullet(M)$ and hence represents a class in
$H^{\mathrm{\ev}}(M,P)$.  The class $ch_P(E) = [ch_P(\nabla,E)]$ is
independent of the choice of module connections
$\nabla_0$ and $\nabla_1$.
\end{proposition}

We note that it is essential to consider $\Z_2$-graded
modules $E$ to define the forms $ch_P(\nabla,E)$ as we
need to be able to consider differences of connections
in order to take traces.  It is straightforward to show
that the assignment $E\mapsto [ch_P(\nabla,E)]$ is
additive with respect to direct sums; i.e. $[ch_P(\nabla\oplus
\nabla',E\oplus E')] = [ch_P(\nabla,E)]+[ch_P(\nabla',E')]$.
We want to show here that the homomorphism $ch_P$
respects the $\tilde{K}^0(M)$-module structure of
$\tilde{K}^0(M,P)$.  We first recall the action of
$\tilde{K}^0(M)$ on $\tilde{K}^0(M,P)$:  if $F=F^+\oplus
F^-$ is a $\Z_2$-graded $\GLtr$-vector bundle on $M$
representing a class in $\tilde{K}^0(M)$ and $E=E^+\oplus
E^-$ is a $\Z_2$-graded $\GLtr$-bundle gerbe module on $P$ for the
lifting bundle gerbe $L\to P^{[2]}$ then $E\cdot F$ is the
$\Z_2$-graded bundle gerbe module $E\hat{\otimes}\pi^*F$,
where $\pi\colon P\to M$ is the projection.  $L$ acts
trivially on $\pi^*F$.  We want to show that
$ch_P(E\hat{\otimes}\pi^*F) = ch_P(E)ch(F)$, where
$ch(F)$ is the ordinary Chern character form for $F$.

Choose a connection $\nabla_E = \nabla_E^+\oplus \nabla_E^-$
on the $\Z_2$-graded module $E=E^+\oplus E^-$ such that
the connections $\nabla_E^+$ and $\nabla_E^-$ are module
connections on $E^+$ and $E^-$ such that the difference
$\nabla_E^+ - \nabla_E^-$ is trace class.  Choose also a
connection $\nabla_F = \nabla_F^+\oplus \nabla_F^-$
on the $\Z_2$-graded $\GLtr$-vector bundle $F$.  Then a
differential form representing $ch_P(E)ch(F)$ is given
by $\exp(f)\tr(\exp(F_{E^+}) - \exp(F_{E^-}))\pi^*\tr(\exp(F_{F^+})
- \exp(F_{F^-}))$.  
This form is equal to
\begin{multline}
\exp(f)\tr(\exp(F_{E^+}\otimes I + I\otimes \pi^*
F_{F^+}) + \exp(F_{E^-}\otimes I + I\otimes \pi^*
F_{F^-})                                              \\
- \exp(F_{E^+}\otimes I + I\otimes \pi^*
F_{F^-}) - \exp(F_{E^-}\otimes I + I\otimes \pi^*
F_{F^+})).
\end{multline}
The connections $(\nabla_{E^+}\otimes I + I\otimes
\pi^*\nabla_{F^+})\oplus (\nabla_{E^-}\otimes I +
I\otimes \pi^*\nabla_{F^-})$ and $(\nabla_{E^+}\otimes
I + I\otimes \pi^*\nabla_{F^-})\oplus (\nabla_{E^-}
\otimes I + I\otimes \pi^*\nabla_{F^+})$ are module
connections, however their difference is not
trace class.  We next choose module connections
$\nabla_{E^+\otimes \pi^*F^+}\oplus \nabla_{E^-\otimes
\pi^*F^-}$ and $\nabla_{E^+\otimes \pi^*F^-}\oplus
\nabla_{E^-\otimes \pi^*F^+}$ on the modules $E^+\otimes
\pi^*F^+\oplus E^-\otimes \pi^*F^-$ and $E^+\otimes \pi^*F^-
\oplus E^-\otimes \pi^*F^+$ respectively.  We choose these
connections so that all of the differences
$\nabla_{E^+\otimes \pi^*F^+} - \nabla_{E^-\otimes \pi^*F^-}$,
$\nabla_{E^+\otimes \pi^*F^+} - \nabla_{E^+\otimes \pi^*F^-}$
and $\nabla_{E^+\otimes \pi^*F^+}- \nabla_{E^-\otimes \pi^*F^+}$
are trace class.  An argument similar to the one above 
shows that one can always
this.  We next define a family of module connections 
$\nabla_{E^{\pm}\otimes \pi^*F^{\pm}}(t)$ by  
\begin{equation}
\nabla_{E^{\pm}\otimes \pi^*F^{\pm}}(t) = t(\nabla_{E^{\pm}}
\otimes I + I\otimes \pi^*\nabla_{F^{\pm}}) + (1-t)
\nabla_{E^{\pm}\otimes \pi^*F^{\pm}}.
\end{equation}
Using the fact that the  
difference of any two of the
connections $\nabla_{E^{\pm}\otimes \pi^*F^{\pm}}$,
or the difference $\nabla_{E^+} - \nabla_{E^-}$ or
$\nabla_{F^+} - \nabla_{F^-}$ is trace class one can show 
that the form $F_{E^+\otimes \pi^*F^+}(t)^{k-1} + 
F_{E^-\otimes \pi^*F^-}(t)^{k-1} - F_{E^+\otimes \pi^*F^-}(t)^{k-1} 
- F_{E^-\otimes \pi^*F^+}(t)^{k-1}$ takes trace 
class values.  Similarly one can show that 
$\dot{A}_{E^+\otimes \pi^*F^+}(t)  + \dot{A}_{E^-\otimes 
\pi^*F^-}(t) - \dot{A}_{E^+\otimes \pi^*F^-}(t) - 
\dot{A}_{E^-\otimes \pi^*F^+}(t)$ takes trace class values.  
>From these two results it is easy to deduce that the form   
\begin{multline}
\label{eq:twisted transgression form}
\dot{A}_{E^+\otimes \pi^*F^+}(t)(F_{E^+\otimes \pi^*F^+}(t))^{k-1}
+ \dot{A}_{E^-\otimes \pi^*F^-}(t)(F_{E^-\otimes \pi^*F^-}(t))^{k-1} \\
- \dot{A}_{E^+\otimes \pi^*F^-}(t)(F_{E^+\otimes \pi^*F^-}(t))^{k-1}
- \dot{A}_{E^-\otimes \pi^*F^+}(t)(F_{E^-\otimes \pi^*F^+}(t))^{k-1}.
\end{multline}
takes trace class values.  
It is also straightforward to check that the trace of the 
form~(\ref{eq:twisted transgression form}) above
descends to $M$.  An argument similar to the
proof that the class $[ch_P(E,\nabla)]$ is independent
of the choice of connection in \cite{bcmms}
shows that the trace of the form~(\ref{eq:twisted
transgression form}) is a transgression form.

We propose that $ch_P(E)$ represents the twisted Chern character.
We summarise the discussion of this section in the following
Proposition.
\begin{proposition}
\label{prop:chern character (even case)}
Let $P$ be a
principal $PU(\cH)$ bundle on $M$ with Dixmier-Douady class $\d(P)$.
The homomorphism $ch_P\colon \tilde{K}^0(M;P) \to
H^{\mathrm{ev}}(M,P)$ satisfies the following properties:
\begin{enumerate}
\item $ch_P$ is natural with respect to pullbacks by maps
$f\colon N\to M$.
\item $ch_P$ respects the $\tilde{K}^0(M)$-module structure of
$\tilde{K}^0(M;P)$.
\item $ch_P$ reduces to the ordinary Chern character in the
case where $P$ is trivial.
\end{enumerate}
\end{proposition}

\section{The Odd Chern Character: Twisted Case}
\label{sec:six}

The relevance of $K^1$ to Type IIA string theory, at least in the
case where there is no background field, has been pointed out by
Witten \cite{Wit}, see also \cite{Hor}.  It turns out
that odd $K$-theory, $K^1(M)$, with
appropriate compact  support conditions, classifies $D$-brane charges in
type IIA  string theory.  Related work also appears in
\cite{HarMoo} and \cite{MatSin}.

Suppose $L\to Y^{[2]}$ is a bundle gerbe with
bundle gerbe connection $\nabla_L$.
Recall that a module connection $\nabla_E$
on a bundle gerbe module $E$ for $L$ is a connection on
the vector bundle $E$ which is compatible with the
bundle gerbe connection $\nabla_L$, i.e. under the
isomorphism $\psi:\pi_1^*E\otimes L\to \pi_2^*E$
the tensor product connection $\pi_1^*\nabla_E\otimes
\nabla_L$ on $\pi_1^*E\otimes L$ is mapped into
the connection $\pi_2^*\nabla_E$ on $\pi_2^*E$.
Suppose now that $L\to P^{[2]}$ is the lifting bundle
gerbe for the principal $PU(\mathcal{H})$ bundle
$P\to M$ and that $\nabla_L$ is a bundle gerbe
connection on $L$ with curvature $F_L$ and curving $f$,
where $F_L = \pi_1^* f - \pi_2^* f$
such that the  associated $3$-curvature (which represents the image, in
real cohomology, of the Dixmier-Douady class of $L$) is
equal to the closed, integral $3$-form $H$.  If $E$ is a trivial
$\Utr$ bundle gerbe module for $L$ then we can consider
module connections $\nabla_E = d + A_E$ on $E$;
however the algebra valued $2$-form
$F_E +fI$ cannot take  trace class values (here $F_E$ denotes the curvature
of the connection $\nabla_E$).
Let $\phi : E \to E$ be an automorphism of the trivial
$\Utr$ bundle gerbe module $E$ that respects
the $\Utr$ bundle gerbe module structure, that is, $\phi
\in \Utr (E)$, then $\phi^{-1} \nabla_E \phi$ is another
module connection for $E$.
Then the difference of connections
$\phi^{-1} \nabla_E \phi - \nabla_E = A(\phi)$ is
a one form on $P$ with values in the trace class
class endomorphisms of $E$.

Recall that the the transformation equation
satisfied by the module connections $\nabla_E$  and $\phi^{-1} \nabla_E
\phi$ is
$$
\pi_1^*\nabla_E + \nabla_L = \psi^{-1}\circ\pi_2^*\nabla_E \circ \psi
$$
and
$$
\pi_1^*\phi^{-1} \nabla_E \phi + \nabla_L =
\psi^{-1}\circ\pi_2^*\phi^{-1} \nabla_E \phi \circ \psi.
$$
Therefore one has the following equality
of $\End(\pi_1^*E\otimes L) = \End(\pi_2^*E)$
valued $1$-forms on $P^{[2]}$:
\begin{equation}\label{inv1}
\pi_1^*A(\phi) = \psi^{-1}\circ\pi_2^*A(\phi) \circ \psi
\end{equation}
Recall also that the curvature satisfies the following equality
of $\End(\pi_1^*E\otimes L) = \End(\pi_2^*E)$
valued $2$-forms on $P^{[2]}$:
\begin{equation}\label{inv2}
\pi_1^*(F_{E} + f I) = \psi^{-1}\circ \pi_2^*(F_{E} + f I)\circ \psi.
\end{equation}
and
\begin{equation}\label{inv3}
\pi_1^*(\phi^{-1} F_E \phi  + f I) = \psi^{-1}\circ \pi_2^*(\phi^{-1} F_E
\phi  + f I)\circ
\psi.
\end{equation}

It follows that the differences $(\phi^{-1} F_E \phi + fI) -
(F_E +fI)$  and hence $(\phi^{-1} F_E \phi + fI)^k - (F_E
+fI)^k$ are differential forms with values in the trace class
endomorphisms of $E$.  Therefore
$\;\tr((\phi^{-1} F_E \phi + fI)^k - (F_E
+fI)^k) $ is well defined and is equal to zero on $P$.  In fact,
by (\ref{inv2}) and (\ref{inv3}), these differential forms descend
to $M$ and are zero there.

In particular, one has
\begin{equation}
\label{eq:character form}
\tr(\exp(\phi^{-1} F_E \phi + fI) - \exp(F_E + fI))
= \exp(f)\tr(\exp(\phi^{-1} F_E \phi) - \exp(F_E)) = 0.
\end{equation}

Let $\nabla_E(s) = s\phi^{-1} \nabla_E \phi + (1-s)\nabla_E$,
where $s\in [0,1]$, denote the linear path of connections
joining $\nabla_E $ and $\phi^{-1} \nabla_E \phi$, and
$F_E(s)$ denote its curvature.
Since $\exp(F_E(s)) - \exp(F_E(0))$ is a differential form 
on $P$ with values in the trace class
class endomorphisms of $E$ for all $s\in [0,1]$, then 
\begin{equation}
\label{trans form}
\partial_s \tr( \exp(F_E(s)) - \exp(F_E(0))) = \tr( \partial_s \exp(F_E(s))) =
d\left( \tr(A(\phi)\exp( F_E(s)))\right)
\end{equation}
 is well defined since
$A(\phi) = \partial_s\nabla_E(s)$ is a one form on $P$ with values in the trace class
class endomorphisms of $E$. 
Integrating (\ref{trans form})  and using (\ref{eq:character form}),
yields the transgression  formula, 
\begin{equation}
\label{eq:transgression formula}
\tr\left(\exp(\phi^{-1} F_E \phi + fI) - \exp(F_E + fI)\right)
= \exp(f)\;d\left(\int_0^1 ds \;\tr(A(\phi)\exp( F_E(s))
)\right).
\end{equation}
By (\ref{eq:character form}), it follows that
$
\displaystyle\int_0^1 ds \;\tr(A(\phi)\exp( F_E(s)))
$
is a closed form. Therefore $\;\exp(f) \displaystyle\int_0^1 ds
\;\tr(A(\phi)\exp( F_E(s)) ) = \;\displaystyle\int_0^1 ds
\;\tr(A(\phi)\exp( F_E(s) + fI) )$ is closed with respect to the
twisted differential
$d-H$. In fact,
by (\ref{inv1}), (\ref{inv2}) and (\ref{inv3}), the differential form
$\;\displaystyle\int_0^1 ds
\;\tr(A(\phi)\exp( F_E(s) + fI) )$
descends  to $M$.

By (\ref{eq:character form}) and (\ref{eq:transgression formula})
we have the following Proposition, except for the last claim.

\begin{proposition}
Suppose that $E$ is $\Utr$ bundle gerbe modules for
the lifting bundle gerbe $L\to P^{[2]}$ equipped with a
bundle gerbe connection $\nabla_L$ and curving $f$, such
that the associated $3$-curvature is $H$.
Suppose that $\nabla_E$ is a module
connection on $E$ and $\phi$ is an automorphism
of $E$ such that $\phi - I_E$ is a fibre-wise trace class
operator. Then the
difference $\phi^{-1} \nabla_E \phi - \nabla_E = A(\phi)$ is
trace class.    Let $ch_P(\nabla_E,\phi) \in
\Omega^{\bullet}(M)$ denote the differential  form on $M$ whose
lift to $P$ is given by $\;\displaystyle\int_0^1 ds
\;\tr(A(\phi)\exp( F_E(s) + fI) )$, where
$\nabla_E(s) = s\phi^{-1} \nabla_E \phi + (1-s)\nabla_E$,
for $s\in [0,1]$, denotes the linear path of connections
joining $\nabla_E $ and $\phi^{-1} \nabla_E \phi$, and
$F_E(s)$ denotes its curvature.
Then
$\;ch_P(\nabla_E,\phi)\;$ is closed with respect to the twisted
differential
$d-H$ on $\Omega^{\bullet}(M)$ and hence represents a class
$[ch_P(\nabla_E,\phi)]$ in $H^{\mathrm{\odd}}(M,P)$.  

Furthermore,
the class $[ch_P(\nabla_E,\phi)]$ is independent of the choice of 
module connection $\nabla_E$ on $E$ and choice of automorphism
$\phi$ of $E$ such that $\phi - I_E$ is a fibre-wise trace class
operator, and is called the odd twisted Chern character of
$[E, \phi] \in K^1(M, P)$, denoted by $ch_P([E, \phi]) \in H^{\mathrm{\odd}}(M,P)$. 
\end{proposition}

\begin{proof}
We have already seen that $ch_P(\nabla_E,\phi)$ is a $d-H$
closed form on $M$. It remains to prove that the twisted cohomology class $[ch_P(\nabla_E,\phi)]$ is independent of the various choices.

\newcommand{\tnabla}{{\widetilde\nabla}}

Let $\nabla_E(s, t)$ be a two parameter family of module connections 
on $E$. Suppose that $\nabla_E(s,t) = d + A(s,t)$ and that 
$F_E(s,t)$ is the curvature of $\nabla_E(s,t)$.   
In the situations that we consider, $\partial_s(A)$ and 
$\partial_s  \exp(F_E))$ are differential forms 
on $P$ with values in the trace class
class endomorphisms of $E$ for all $s\in [0,1]$, where we 
have supressed the dependence of $A$ and $F_E$ on $s$ and $t$.  
Let $P(X)$ denote the invariant function  $\tr(\exp(X))$ and 
let $P'$ and $P''$ denote the first and second differentials 
of $P$ respectively.   
An easy calculation shows that 
$\tr(\partial_s(\partial(A)F_E^{k-1}) - 
\partial_t((\partial_s(A)F_E^{k-1})) = 
\tr(\partial_t(A)\partial_s(F_E^{k-1}) - 
\partial_s(A)\partial_t(F_E^{k-1}))$.  
We rewrite the right hand side as follows.  We have 
\begin{eqnarray*} 
&   & \tr(\partial_t(A)\partial_s(F_E^{k-1}) - 
\partial_s(A)\partial_t(F_E^{k-1}))               \\          
& = & \tr\left\{ \sum^{k-1}_1 \partial_t(A)F_E^{i-1}(d\partial_s(A) 
+ \partial_s(A) A + A\partial_s(A))F_E^{k-i-1}  \right.      \\  
&   & \left.- \sum^{k-1}_1 \partial_s(A)F_E^{i-1}
(d\partial_t(A) + \partial_t(A) A 
+ A\partial_t(A))F_E^{k-i-1}  \right\}               \\ 
& = & \tr\left\{ \sum^{k-1}_1 \partial_t(A) 
F_E^{i-1} d\partial_s(A)F_E^{k-i-1} 
- \sum^{k-1}_1 d\partial_t(A)F_E^{k-i-1}\partial_s(A)F_E^{i-1} \right. \\  
&   & \left. + \sum^{k-1}_1 \partial_t(A) 
F_E^{i-1}\partial_s(A)AF_E^{k-i-1} - 
\sum^{k-1}_1 \partial_t(A)AF_E^{k-i-1}\partial_s(A)F_E^{i-1} \right.   \\ 
& & \left.+ \sum^{k-1}_1 \partial_t(A) F_E^{i-1}A\partial_s(A)F_E^{k-i-1} 
- \sum^{k-1}_1 \partial_t(A)F_E^{k-i-1}\partial_s(A)F_E^{i-1}A \right\} 
\end{eqnarray*} 
where we have used the invariance of $\tr$.  Reindexing the sums 
we get 
\begin{multline*} 
\tr\left\{ -\sum^{k-1}_1 d\partial_t(A)F_E^{i-1}\partial_s(A)F_E^{k-i-1} 
+ \sum^{k-1}_1 \partial_t(A)F_E^{i-1}d\partial_s(A)F_E^{k-i-1} \right.   \\  
\left.- \sum^{k-1}_1 \partial_t(A)F_E^{i-1}\partial_s(A)[F_E^{k-i-1},A] 
+ \sum^{k-1}_1\partial_t(A)[F_E^{i-1},A]\partial_s(A)F_E^{k-i-1}\right\} 
\end{multline*}
which equals $-d\tr\sum^{k-1}_1 \partial_t(A)F_E^{i-1}\partial_s(A) 
F_E^{k-i-1}$ on using the Bianchi identity.  To recap, we have 
the following identity: 
\begin{multline} 
\tr(\partial_t(\partial_s(A(s,t))\exp(F_E(s,t))) 
- \partial_s(\partial_t(A(s,t)) 
\exp(F_E(s,t)))) \\ 
 = dP''(F_E(s,t),\partial_t(A(s,t)),\partial_s(A(s,t))).  
\end{multline} 
Next observe that we can write 
$\tr(\partial_s(\partial_t(A(s,t)) \exp(F_E(s,t))))$ 
as 
$$
\partial_s \tr\left(\partial_t(A(s,t)) 
\exp(F_E(s,t)) - \partial_t(A(0, t) \exp(F_E(0,t))\right) 
$$
where the expression inside the trace takes trace class values 
due to our hypotheses on $A(s,t)$ and $F_E(s,t)$.  
Integrating with respect to $s, t$, we obtain the identity,
\begin{equation}\label{trans0}
\displaystyle d \int_0^1 \int_0^1 ds\wedge dt  \; P''(F_E(s,t) , 
\partial_t (A(s, t)), \partial_s (A(s, t))) = 
\end{equation}
\begin{equation}\label{trans1}
\displaystyle\int_0^1 dt  \tr\left(\partial_t(A(1, t)) 
\exp(F_E(1,t)) - \partial_t(A(0, t)) \exp(F_E(0,t))\right)
\end{equation}
\begin{equation}\label{trans2}
-  \int_0^1 ds \; P'(F_E(s,1) , \partial_s (A(s, 1)))
+ \int_0^1 ds \; P'(F_E(s,0) , \partial_s (A(s, 0)))
\end{equation}
The first 2-parameter family that we will consider is the following.
$\nabla_E(s, t) = (1-s) (d + A(t)) + s \phi^{-1} (d + A(t))\phi$,
where $A(t) =  tA_1 + (1-t)A_0$ and $A_0, A_1$ are connection 1-forms on $E$.  
To prove the independence of the choice
of the connection 1-form, from (\ref{trans0}), (\ref{trans1}), (\ref{trans2}),
it suffices to show that (\ref{trans1}) vanishes.
 We compute,
$  \partial_t \nabla_E(s, t) = - (1-s) (A_1-A_0) +  s\phi^{-1}(A_1-A_0) \phi$. 
In particular, $ \partial_t \nabla_E(1, t) = \phi^{-1}(A_1-A_0) \phi$
and $ \partial_t \nabla_E(0, t) = (A_1-A_0) $. Also, $F_E(1,t)
= \phi^{-1} F_E(0,t)\phi$. 
So (\ref{trans1}) becomes,
$$
\displaystyle \int_0^1 dt  \tr\left( \phi^{-1}(A_1-A_0) 
\phi\exp(\phi^{-1} F_E(0,t)\phi) - (A_1-A_0)\exp(F_E(0,t))\right)
= 0
$$
by the invariance of $P$. Therefore 
$$
ch_P(\nabla_E(1), \phi) - ch_P(\nabla_E(0), \phi) = 
$$
$$
\displaystyle (d - H) \exp(f) \int_0^1 \int_0^1 ds\wedge dt  
\; P''(F_E(s,t) , \partial_t \nabla_E(s, t), \partial_s \nabla_E(s, t)),  
$$
as claimed, where $\nabla_E(j) = d + A_j, \; j=0,1.$

To prove the independence of the choice of $\phi$, we choose the 2-parameter 
family given by $\nabla_E(s, t) = (1-s) (d + A) + s \phi_t^{-1} (d + A)\phi_t$. It suffices to 
show that (\ref{trans1}) is an exact form. We compute, 
$  \partial_t \nabla_E(s, t) =-  s \phi_t^{-1}\dot\phi_t\phi_t^{-1}(d+A) 
\phi_t +  s\phi_t^{-1} (d + A)\dot\phi_t$ is a trace class operator. 
In particular, $ \partial_t \nabla_E(1, t) = -  \phi_t^{-1}\dot\phi_t\phi_t^{-1}(d+A) \phi_t +  \phi_t^{-1} (d + A)\dot\phi_t$. 
and $ \partial_t \nabla_E(0, t) = 0 $. Set $\nabla_E = d+A$ and  let $F_E$
denote its curvature.
Using the fact that $P(X) = \tr(\exp(X))$, we get that (\ref{trans1}) becomes
$$
\tr\left(d(\dot\phi_t \phi_t^{-1}) \exp(F_E)) + 
\dot\phi_t \phi_t^{-1}[\exp(F_E)) , A]\right) 
$$
$$
= \tr\left(d(\dot\phi_t \phi_t^{-1} \exp(F_E)) - 
\dot\phi_t \phi_t^{-1} d(\exp(F_E)) )
+ \dot\phi_t \phi_t^{-1}[\exp(F_E) , A]\right) 
$$
$$
= d\tr\left(\dot\phi_t \phi_t^{-1} \exp(F_E)\right),
$$
since the other terms vanish by repeated application of the Bianchi identity
$dF_E - [F_E, \nabla_E] = 0$. Observe that $\dot\phi_t \phi_t^{-1}$
is a trace class operator, so that all the terms are of trace class. 
Therefore 
$$
ch_P(\nabla_E, \phi_1) - ch_P(\nabla_E, \phi_0) = 
$$
$$
\displaystyle (d - H) \exp(f) \int_0^1 \int_0^1 ds\wedge dt  \; P''(F_E(s,t) , \partial_t \nabla_E(s, t), \partial_s \nabla_E(s, t)) 
$$
$$
+ (d - H)  \exp(f) \int_0^1 dt \;    \tr\left(\dot\phi_t \phi_t^{-1} \exp(F_E)\right),  
$$
as claimed.

\end{proof}

Let $\text{Mod}_{\Utr}^1(L)$ denote the
semi-group of all pairs $(E, \phi)$ consisting of a trivial
$\Utr$ bundle gerbe module $E$ for
the lifting bundle gerbe $L\to P^{[2]}$ associated to
a principal $PU(\mathcal{H})$ bundle $P\to M$ with
Dixmier-Douady class equal to $\delta(P)$, together
with an automorphism $\phi: E \to E$
such that $\phi - I_E$ is a fibre-wise trace class
operator.
Then we have defined a map
$ch_{P}\colon \text{Mod}_{\Utr}^1(M,L)\to H^{\mathrm{odd}}(M,P)$.
Observe that $\text{Mod}_{\Utr}^1(L) = [P, \Utr]^{PU}$, and
by Palais theorem, we have the natural isomorphism
$\text{Mod}_{\Utr}^1(M,L) =  {K}^1(M,P)$ that is induced by
the inclusion $\Utr\subset U_K$. Therefore  we
get a map
$$
ch_{P}\colon {K}^1(M,P)\to
H^{\mathrm{odd}}(M,P).
$$
We assert that this map
defines the odd Chern character for twisted
$K$-theory.  It can be shown that the odd
Chern character for twisted $K$-theory
is uniquely characterised by requiring that
it is a functorial homomorphism which is compatible
with the ${K}^0(M)$-module
structure on ${K}^1(M,P)$ and reduces to
the ordinary odd Chern character when $P$ is trivial.
It is easy to check that the map $ch_{P}
\colon {K}^1(M,P)\to H^{\mathrm{odd}}(M, P)$
is functorial with respect to smooth maps
$f\colon N\to M$. The proof that $ch_{P}$ is a
homomorphism is exactly as in the even degree case,
cf. section 9, \cite{bcmms}.
The proof
that $ch_{P}$ is compatible with the $\tilde{K}^0(M)$-module
structure of ${K}^1(M,P)$ is exactly as in the earlier section for
twisted $K^0$.

\section{Equivariant Chern Character: Twisted case}
\label{sec:seven}

Suppose that $G$ is a compact Lie group acting
on the smooth manifold $M$.  We want to define
a Chern character for the equivariant twisted
$K$-theory $\tilde{K}^0_G(M,[H])$.  In this
equivariant setting the twisting is done by a
class in $H^3_G(M,\Z)$, where $H_G$ denotes
equivariant cohomology.  Recall that $H_G$ is
defined by first constructing the Borel space $M^G =
M\times_G EG$ and then
setting $H^\bullet_G(M,\Z) = H^\bullet(M^G,\Z)$.
When $G$ acts freely on $M$ one can show that there
is an isomorphism $H^\bullet_G(M,\Z) = H^\bullet(M/G,\Z)$.
Note that equivariant cohomology groups are a lot larger
than ordinary cohomology groups; $H^\bullet_G(M,\Z)$ is a
module over $H^\bullet_G(pt,\Z) = H^\bullet(BG,\Z)$.

A $G$-equivariant principal $PU$-bundle $P$ determines an 
element of $H^3_G(M,\Z)$. 
To see this, we note that $P$ induces a principal 
$PU$-bundle $P^G$ over $M^G$ by first lifting it to the 
product $M \times EG$ and by equivariance, it descends 
to $M^G$. So by standard Dixmier-Douady theory, it determines  
a class in $H^3(M^G,\Z) = H^3_G(M,\Z)$.
Brylinski \cite{Bry} 
identifies $H^3_G(M,\Z)$ with equivalence classes of
$G$-equivariant gerbes on $M$, and Meinrenken 
\cite{Mein} identifies $H^3_G(M;\Z)$ with equivalence 
classes of $G$-equivariant bundle gerbes on $M$.  
$G$-equivariant principal $PU$ bundles on $M$ form a 
subgroup of $H^3_G(M:\Z)$.  
In the case when $M=G$ and $G$ acts by conjugation 
on itself, the principal $PU$-bundles that are associated 
bundles to  the canonical loop group $LG$ bundle 
over $G$ via positive energy representations of the loop group $LG$,
are all $G$-equivariant in our sense. Although these representations
are strongly continuous, they are equivalent to norm continuous ones.

We now explain how a $G$-equivariant $PU$ bundle
on $M$ gives rise to a $G$-equiavriant bundle
gerbe.  Suppose $P$ is a $G$-equivariant $PU$ bundle
on $M$, by replacing $M$ and $P$ by $M\times EG$
and $P\times EG$ respectively, we may assume without
loss of generality that
$G$ acts freely.  Associated to the $PU$-bundle
$P/G$ on $M/G$ is the lifting bundle gerbe
$J\to (P/G)^{[2]} = P^{[2]}/G$.  The projection
$P\to P/G$ covering $M\to M/G$ induces a map
$P^{[2]}\to (P/G)^{[2]}$.  Let $L$ denote the pullback
of $J$ to $P^{[2]}$ under this map.  Then $L$ is a
$G$-equivariant line bundle on $P^{[2]}$.  It is easy
to see that in fact $L\to P^{[2]}$ has a bundle gerbe
product compatible with the isomorphisms $\sigma_g\colon
L_{(p_1,p_2)}\to L_{g(p_1,p_2)}$
of~(\ref{eq:G-equivariant line bundle}).

Given a $G$-equivariant $PU$-bundle 
$P$ with Dixmier-Douady class $\delta(P)$ in $H^3_G(M,\Z)$ we define
the equivariant twisted $K$-theory $\tilde{K}^0_G(M, P)$ as follows.
We let $\cH_G$ be a separable $G$-Hilbert space
in which every irreducible representation of
$G$ occurs with infinite multiplicity.  We then
define $\tilde{K}^0_G(M, P)$ to be the space of
$G$-homotopy classes of $G$-equivariant,
$PU$-equivariant maps $P\to BGL_{\K}$, where $P\to M$
is a $G$-equivariant principal $PU$ bundle on $M$
corresponding to the class $\delta(P)$ in $H^3_G(M,\Z)$.
Here $BGL_{\K}$ is the quotient $GL(\cH_G)/GL_{\K}(\cH_G)$.
The odd-dimensional group $\tilde{K}^1_G(M, P)$ is
defined to be the space of $G$-homotopy classes of $G$-equivariant,
$PU$-equivariant maps $P\to GL_{\K}$.  Again $GL_{\K}$ is
$GL_{\K}(\cH_G)$.
Again there are various other equivalent definitions
of $\tilde{K}^0_G(M, P)$; we have in analogy with
Proposition~\ref{prop:various twisted K-theory definitions}

\begin{proposition}
\label{prop:various definitions of equivariant twisted K-theory}
Given a $G$-equivariant $PU$ bundle $P$ on $M$
corresponding to a class in $H^3_G(M,\Z)$, we have
the following equivalent definitions of $\tilde{K}^0_G(M, P)$:
\begin{enumerate}
\item space of $G$-homotopy classes of $G$-equivariant
sections of the associated bundle $P\times_{PU}BGL_{\K}$.
\item $G$-isomorphism classes of $G$-equivariant, $PU$ covariant
$GL_{\K}$ bundles on $P$.
\end{enumerate}
\end{proposition}

Associated to the $G$-equivariant $PU$ bundle $P\to M$
is the $G$-equivariant lifting bundle gerbe $L\to P^{[2]}$.
We can formulate the notion
of a $G$-equivariant bundle gerbe module for $L$: this
is $G$-equivariant $GL_{\K}$-vector bundle $E\to P$ on
which $L$ acts via a $G$-equivariant isomorphism
$\pi_1^*E\otimes L\to \pi_2^*E$.  Again we require
an extra condition relating to the action of $U(\cH)$ on
the principal $GL_{\K}$ bundle associated to $E$.
As before we form the principal $GL(\cH)$ bundle $GL(E)$
on $P$ with fibre at $p$ equal to the isomorphisms
$\cH \to E_p$.  This is a $G$-equivariant $GL(\cH)$ bundle.
The $GL_{\K}$-structure on $E$ induces a reduction $R$ of
$GL(E)$ to a $GL_{\K}$-bundle.  Again $R$ is a $G$-equivariant
$GL_{\K}$ bundle.  We require that the action of $U(\cH)$ on
$GL(E)$ given by sending $f$ to $ufu^{-1}$ is a $G$-map
preserving $R$.
The set of $G$-equivariant
bundle gerbe modules for $L$ forms a semi-group
$\Mod_{GL_{\K}}(L,M)_G$ under
direct sum.  In analogy with Proposition~\ref{prop:bg K-theory
= twisted K-theory} we have the following result.
\begin{proposition}
\label{prop:equiv bg K-theory = equiv twisted K-theory}
If $L\to P^{[2]}$ is the $G$-equivariant lifting bundle
gerbe associated to the $G$-equivariant $PU$ bundle $P\to M$
corresponding to the class $\delta(P)\in H^3_G(M,\Z)$ then
$$
\tilde{K}^0_G(M, P) = \Mod_{GL_{\K}}(L,M)_G.
$$
\end{proposition}
Again the result remains valid if we replace $G$-equivariant
$GL_{\K}$-vector bundles by $G$-equivariant $\GLtr$-vector bundles.
We must perform this replacement in order to define the
Chern character.

Before we define the Chern character we must first
have a model for an equivariant de Rham theory.
Good references for this are \cite{BerGetVer} and \cite{MatQui}.  We shall
use the Cartan model to compute
equivariant de Rham theory following \cite{BerGetVer}.
Recall that one introduces the algebra $S(\mathfrak{g}^*)
\otimes \Omega^\bullet (M)$.  This is given a $\Z$-grading
by $\deg (P\otimes \omega) = 2\deg (P) + \deg(\omega)$.
An operator on $S(\mathfrak{g}^*)\otimes \Omega^\bullet(M)$ is
defined by the formula $d_{\mathfrak{g}} = d - i$, where $i$
denotes contraction with the vector field on $M$ induced
by the infinitesimal action of an element of $\mathfrak{g}$
on $M$.  $d_{\mathfrak{g}}$ preserves the $G$-invariant sub-algebra
$\Omega^\bullet_G(M) = (S(\mathfrak{g}^*)
\otimes \Omega^\bullet(M))^G$ and raises
the total degree of an element of
$\Omega_G^\bullet(M)$ by one.  If we restrict $d_{\mathfrak{g}}$ to
$\Omega_G^\bullet(M)$ then
$d_{\mathfrak{g}}^2 = 0$.  It can be shown that for $G$ compact
the cohomology of the complex
$(\Omega_G^\bullet(M),d_{\mathfrak{g}})$ is equal to the
equivariant cohomology $H^\bullet_G(M)$.

We briefly recall, following \cite{BerGetVer},
the construction of Chern classes living in equivariant
cohomology $H^{\ev}_G(M)$
associated to a $G$-equivariant vector bundle $E\to M$.
Suppose that $\nabla$ is a $G$-invariant connection
on $E$.  By this we mean the following (cf. \cite{Bry}).
The connection $\nabla$ induces by pullback
connections $p_1^*\nabla$ and $m^*\nabla$ on the bundles
$p_1^*E$ and $m^*E$ on $M\times G$.  There exists a
$1$-form $A$ on $M\times G$ with values in the endomorphism
bundle $\End(p_1^*E) = p_1^*\End(E)$ such that
\begin{equation}
\label{eq:moment 1-form}
p_1^*\nabla = \sigma^{-1}m^*\nabla \sigma + A.
\end{equation}
The condition that $\nabla$ is $G$-invariant means that $A$
vanishes in the $M$-direction.
We associate a moment map $\mu\in \Gamma(M,\End(E))\otimes
\mathfrak{g}^*$ to $\nabla$ by defining
\begin{equation}
\label{eq:moment map}
\mu(\xi)(s) = \mathcal{L}_{\xi}(s) - \nabla_{\hat{\xi}}(s)
\end{equation}
where $s$ is a section of $E$ and $\hat{\xi}$ is the
vector field on $M$ induced by the infinitesimal
action of $G$.  Here the Lie derivative on sections is
defined by
$$
\mathcal{L}_{\xi}(s) = \frac{d}{dt}|_{t=0} \exp(t\xi)\cdot s,
$$
where $G$ acts, for $g$ close to $1$, on sections
by the formula $(gs)(m) = gs(g^{-1}m)$.  This last
point perhaps requires some more explanation.  $G$ itself
does not act on $E$ but rather there is the isomorphism
$\sigma\colon p_1^*E\to m^*E$ over $M\times G$.  As we have
explained already, $\sigma$ consists of a family of
isomorphisms $\sigma_g\colon E\to g^*E$ which are
smooth in $g$.  For $g$ close to the identity therefore
we can assume that $g$ acts on $E$.  Note that
$\mu$ actually lives in $(\Gamma(M,\End(E))\otimes \mathfrak{g}^*)^G$.

There is however another way \cite{Bry} to look at the moment map
which will be more useful for our purposes.
Recall the $1$-form $A$ defined
above by comparing the pullback connections
$p_1^*\nabla$ and $m^*\nabla$ via $\sigma$.  One can
show, as a result of the associativity condition on
$\sigma$ that $A$ is left invariant under the left action
of $G$ on $M\times G$ where $G$ acts on itself by left multiplication.
Since $\nabla$ is $G$-invariant $A$
vanishes in the $M$-direction.
Therefore we can define a section $\mu_E\in \Gamma(M,\End(E))\otimes
\mathfrak{g}^*$ by
$$
(\mu_E)_{\xi}(s)(m) = A((m,1);(0,\xi))(s(m)).
$$
Again note that $\mu_E$ belongs to
$(\Gamma(M,\End(E))\otimes \mathfrak{g}^*)^G$.  We
have the equality $\mu = \mu_E$.

To define an equivariant Chern character we
put $ch_G(E,\nabla) = \tr(\exp(F_{\nabla} - \mu))$.
One can show that $ch_G(E,\nabla) \in (S(\mathfrak{g}^*
\otimes \Omega^\bullet(M))^G = \Omega_G^\bullet(M)$
is equivariantly closed and moreover that the class
defined by $ch_G(E,\nabla)$ in $H^{\ev}_G(\Omega_G^\bullet(M))$ is
independent of the choice of connection $\nabla$.

We need to explain how the theory of bundle gerbe connections
and curvings carries over to the equivariant case.  So
suppose that $L\to Y^{[2]}$ is a $G$-equivariant bundle gerbe on
$M$.  So $Y$ has a $G$-action and $\pi\colon Y\to M$
is a submersion 
which is also a $G$-map.  $L\to Y^{[2]}$ is a
$G$-equivariant line bundle with a $G$-equivariant product.
Suppose that $L$ comes equipped with a connection
$\nabla$ that preserves the bundle gerbe product (\ref{eq:bundle
gerbe product}) and is $G$-invariant.  Since $L$ is
$G$-equivariant there is a line bundle isomorphism
$\sigma\colon p_1^*L\to m^*L$ covering the identity
on $Y^{[2]}\times G$ (corresponding to the family of
maps $\sigma_g$ of (\ref{eq:G-equivariant line bundle})).
As above $\nabla$ induces connections $p_1^*\nabla$ and
$m^*\nabla$ on $p_1^* L$ and $m^* L$
respectively and we can define a $1$-form
$A_L$ on $Y^{[2]}\times G$ in the same manner as
equation~(\ref{eq:moment 1-form}) above.
Again it is easy to see that $A_L$ is left invariant under the
action of $G$ on $Y^{[2]}\times G$ where $G$ acts on itself
by left multiplication and since $\nabla$ is $G$-invariant
$A_L$ vanishes in the $Y^{[2]}$-direction.  It is also
straightforward to see that we have the equation
\begin{equation}
(\pi_1\times 1)^*A_L - (\pi_2\times 1)^*A_L +
(\pi_3\times 1)^* A_L = 0
\end{equation}
in $\Omega^1(Y^{[3]}\times G)$ (here for example
$(\pi_1 \times 1)(y_1,y_2,y_3,g) = (y_2,y_3,g)$).
It follows that we can find a $1$-form $B$ on
$Y\times G$ so that
$$
A_L = (\pi_2\times 1)^*B - (\pi_1\times 1)^*B.
$$
One can show that it is possible to choose $B$
so that it is invariant under $G$ and vanishes
in the $Y$-direction.  It follows that we can
define maps $\tilde{\mu}_L\colon Y^{[2]}\to \mathfrak{g}^*$,
$\mu_L\colon Y\to \mathfrak{g}^*$ such that
$\tilde{\mu}_L(y_1,y_2)(\xi) = A((y_1,y_2,1);(0,0,\xi))$
and $\mu_L(y)(\xi) = B((y,1);(0,\xi))$ and moreover
$\tilde{\mu}_L = \delta(\mu_L) = \pi_2^*\mu_L - \pi_1^*\mu_L$.

The degree two element $F-\tilde{\mu}_L$ of
$\Omega_G^\bullet(Y^{[2]})$ is equivariantly closed,
i.e. closed with respect to $d_{\mathfrak{g}}$.  We have
$F-\tilde{\mu}_L = \delta (f-\mu_L)$ and, since $\delta$
commutes with $d_{\mathfrak{g}}$, we have $\delta d_{\mathfrak{g}}
(f-\mu_L) = 0$.  It follows from here that
$d_{\mathfrak{g}}(f-\mu_L) = \pi^*H$ for some degree three
element $H$ of $\Omega_G^\bullet(M)$ which is
necessarily equivariantly closed.

We also need to explain how twisted cohomology
arises in the equivariant case.  Suppose $H$ is the
equivariantly closed degree three element of
$\Omega_G^\bullet(M)$ associated to an invariant bundle
gerbe connection on the equivariant bundle gerbe
associated to the $G$-equivariant $PU$ bundle $P$
on $M$.  Thus the image of $H$ in $H^3(\Omega_G^\bullet(M))$
represents the image of the equivariant Dixmier-Douady
class of $P$ in $H^3_G(M)$.  As in Section~\ref{sec:three},
we introduce a twisted differential on the algebra
$\Omega_G^\bullet(M)$ by $d_H = d_{\mathfrak{g}} - H$.
Because $H$ is equivariantly closed, $d_H^2 = 0$.
We denote the cohomology of the complex $(\Omega_G^\bullet(M),
d_H)$ by $H^\bullet_G(M,H)$.  

To define the twisted equivariant Chern character we need to
explain what an equivariant module connection is.
Suppose firstly that the equivariant bundle gerbe
$L\to P^{[2]}$ has a $G$-invariant connection $\nabla$
and `equivariant curving' $f-\mu_L$.
If $E\to P$ is a $G$-equivariant bundle gerbe module
for the equivariant bundle gerbe $L\to P^{[2]}$ then a
$G$-equivariant module connection for $E$ is a connection
on the vector bundle $E$ which is invariant with respect to
the $G$-action on $E$ and which is compatible with the bundle
gerbe connection $\nabla$ in the sense
of the equation~(\ref{eq:module connection}).  Since the isomorphism
$\pi_1^*E\otimes L\to \pi_2^*E$ is a $G$-isomorphism we
have the commuting diagram
$$
\begin{array}{ccc}
p_1^*(\pi_1^*E\otimes L) & \stackrel{\scriptstyle \pi_1^*\sigma_E
\otimes \sigma_L}{\rightarrow} &
m^*(\pi_1^*E\otimes L)                      \\
\scriptstyle{p_1^*\psi} \downarrow & & \downarrow
\scriptstyle{m^*\psi}              \\
p_1^*\pi_2^*E & \stackrel{\scriptstyle
\pi_2^*\sigma_E}{\rightarrow} &  m^*\pi_2^*E
\end{array}
$$
over $P^{[2]}\times G$.  From this diagram we deduce that
the $\End(E)$-valued $1$-form $A_E$ on $P\times G$ and the 1-form
$A_L$ on $P^{[2]}\times G$ satisfy the equation
\begin{equation}
(\pi_1\times 1)^* A_E\otimes I + I\otimes A_L =
p_1^*\psi^{-1} (\pi_2\times 1)^* A_E p_1^*\psi
\end{equation}
in $\Omega^1(P^{[2]}\times G)\otimes p_1^*\End(\pi_1^*E\otimes L)$.
        From here it easy to conclude that the moment maps
$\mu_E$ and $\delta(\mu_L)$ satisfy
\begin{equation}
\label{eq:descent for moment maps}
\pi_1^*(\mu_E - \mu_L) = \psi^{-1}\pi_2^*(\mu_E - \mu_L)\psi.
\end{equation}
If we are given a $\Z_2$-graded equivariant bundle gerbe
module $E = E_0 \oplus E_1$ we can define a twisted
equivariant Chern character $ch_P^G(E)$ taking values
in the twisted equivariant cohomology group
$H_G(M,H)$ as follows.  We first of all choose a
$\Z_2$-graded equivariant bundle gerbe module connection
$\nabla = \nabla_0\oplus\nabla_1$ on $E$ such that
the difference $\nabla_0 - \nabla_1$ is trace
class when considered in local trivialisations.
It is easily seen
that the difference
\begin{equation}
\exp(F_0 + \mu_{E_0} + f - \mu_L) -
\exp(F_1 + \mu_{E_1} + f - \mu_L)
\end{equation}
is trace class.  It follows from the
equations~(\ref{eq:curvature transfm law})
and~(\ref{eq:descent for moment maps})
that the trace of this expression descends
to $M$.  The resulting form is easily seen
to be closed with respect to the twisted
equivariant differential.

\section{ Chern Character: the twisted holomorphic case}
\label{sec:eight}

In this section, we introduce the theory of holomorphic
bundle gerbes, holomorphic bundle gerbe modules,
and  holomorphic bundle gerbe $K$-theory, as well
as the Chern character in this context. For the possible
relevance of the material in this section to $D$-brane
theory, cf. \cite{Sh}, \cite{KaOr}.

\subsection{Holomorphic bundle gerbes.} Holomorphic
gerbes have been studied by various authors such as
\cite{Bry} using categories, stacks and coherent sheaves. Our goal in this
subsection is to sketch a holomorphic bundle theory analogue,
as accomplished by Murray \cite{Mur} in the smooth case.

In this section $M$ denotes a complex manifold.
A holomorphic bundle gerbe on $M$ consists of a
holomorphic submersion $\pi\colon Y\to M$, together 
with a holomorphic line bundle $L\to Y^{[2]}$
on the fibre product $Y^{[2]} = Y\times_{\pi}Y$.
$L$ is required to have a product, that is a holomorphic
bundle isomorphism which on the fibres takes the
form
\begin{equation}
\label{eq:hol bundle gerbe product}
L_{(y_1,y_2)}\otimes L_{(y_2,y_3)}\to L_{(y_1,y_3)}
\end{equation}
for points $y_1$, $y_2$ and $y_3$ all belonging to the
same fibre of $Y$.  This product is required to be
associative in the obvious sense.  If one chooses
an open cover $\U = \{U_i\}$ of $M$ over which
there exist local holomorphic sections $s_i\colon U_i\to Y$ of
$\pi$, then we can form holomorphic maps $(s_i,s_j)\colon
U_{ij}\to Y^{[2]}$ and pull back the holomorphic line bundle
$L$ to get a family of holomorphic line bundles $L_{ij}$ on
$U_{ij}$ together with holomorphic isomorphisms $L_{ij}\otimes
L_{jk}\to L_{ik}$.  One example of such a structure is the
case when we have a holomorphic principal $PGL$ bundle $Y$
over $M$.  That is, $Y$ has holomorphic local trivialisations.
Then we can form the lifting bundle gerbe $L\to Y^{[2]}$ in the
standard manner.  $L$ is a holomorphic line bundle and the
bundle gerbe product on $L$ induced from the product in the
group $GL$ is holomorphic.

Every holomorphic bundle gerbe on $M$ gives rise to a class in
the sheaf cohomology group $H^2(M,\underline{\cO}^*_M)$,
where $\underline{\cO}^*_M$ denotes sheaf of nonvanishing
holomorphic functions on $M$.  We
associate a $\cO^*$-valued \v{C}ech $2$-cocycle $\epsilon_{ijk}$ to
the holomorphic bundle gerbe $L\to Y^{[2]}$ in the manner
described in \cite{Mur}.  We first pick an open cover $\{U_i\}$ of
$M$ such that there exist holomorphic local sections
$s_i\colon U_i\to Y$ of $\pi\colon Y\to M$.  Then we can form the
holomorphic maps $(s_i,s_j)\colon U_{ij}\to Y^{[2]}$ and form
the pullback line bundle $L_{ij} = (s_i,s_j)^*L$.  $L_{ij}$ is a
holomorphic line bundle on $U_{ij}$.  We then pick holomorphic
sections $\sigma_{ij}\colon U_{ij}\to L_{ij}$ and define a
holomorphic function $\epsilon_{ijk}\colon U_{ijk}\to \cO^*$.
It is easy to see that $\epsilon_{ijk}$ satisfies the \v{C}ech
$2$-cocycle condition $\delta(\epsilon_{ijk}) = 1$.  The
class in $H^2(M,\underline{\cO}^*_M)$ determined by $\epsilon_{ijk}$
can be shown to be independent of all the choices.  Brylinski
\cite{Bry} proves that there is an isomorphism between
equivalence classes of holomorphic gerbes on $M$ and
$H^2(M,\underline{\cO}^*_M)$.  To obtain an analogue of this
result for holomorphic bundle gerbes we would need to introduce the
notion of stable isomorphism of holomorphic bundle gerbes,
analogous to what was done in
\cite{MurSte} in the smooth case.

Consider now holomorphic bundle gerbes
$L\to Y^{[2]}$ with a given hermitian metric,
and call these hermitian holomorphic bundle gerbes.
A compatible bundle gerbe connection $\nabla$ on a
hermitian holomorphic bundle gerbe
$L\to Y^{[2]}$ is a connection on the line bundle
$L$ which preserves a hermitian metric and
is compatible with the product
given in (\ref{eq:hol bundle  gerbe product}), i.e. $\nabla(st) =
\nabla(s)t +  s\nabla(t)$ for sections $s$ and $t$ of $L$.
In \cite{Mur} it is shown that bundle gerbe
connections that preserve a hermitian metric always exist in the smooth
case, and  this implies the existence in the holomorphic
case.  The curvature
$F_{\nabla} \in \Omega^{1, 1}(Y^{[2]})\subset \Omega^2(Y^{[2]})$ of a
compatible bundle gerbe connection
$\nabla$ is  easily seen to satisfy $\delta(F_{\nabla}) = 0$ in
$\Omega^2(Y^{[3]})$.  It follows that we can find a
$(1, 1)$-form $f$ on $Y$ such that $F_{\nabla} =
\delta(f) =
\pi_2^*f - \pi_1^*f$.  $f$ is unique up to $(1,1)$-forms
pulled back from $M$.  A choice of $f$ is called a
choice of a curving for $\nabla$.  Since $F_{\nabla}$ is
closed we must have $df = \pi^*\omega$ for some necessarily
closed $3$-form $\omega\in \Omega^{2, 1}(M)\oplus \Omega^{1, 2}(M)$ on
$M$.  $\omega$ is called  the $3$-curvature of the compatible bundle gerbe
connection $\nabla$  and curving $f$.  It follows from \cite{Mur} that
$\omega$ has integral periods, and the Dixmier-Douady
invariant of the hermitian holomorphic bundle gerbe
$L\to Y^{[2]}$ is $[\omega]\in (H^{1,2}(M, \C)\oplus H^{2, 1}(M, \C))\cap 
H^3(M, \Z)
\subset H^3(M, \Z)$. This says in particular that not all classes
in $H^3(M, \Z)$ arise as the Dixmier-Douady invariant of
hermitian holomorphic bundle gerbes.

If we  instead merely
considered holomorphic bundle gerbes $L\to Y^{[2]}$, then a
compatible bundle gerbe connection $\nabla$ would have
curvature $F_{\nabla} \in \Omega^{2, 0}(Y^{[2]})\oplus \Omega^{1,
1}(Y^{[2]})\subset\Omega^2(Y^{[2]})$. Arguing as above,
      we can find a choice of curving for $\nabla$  which is a
form $f$ on $Y$ that is in the subspace $\Omega^{2, 0}(Y)\oplus
\Omega^{1, 1}(Y)\subset\Omega^2(Y)$, satisfying $F_{\nabla} =
\delta(f) =\pi_2^*f - \pi_1^*f$. It follows as before that
if $\omega$ is the $3$-curvature of the compatible bundle gerbe
connection $\nabla$  and curving $f$, then $\omega$  is a closed
form in 
$\Omega^{3, 0}(M)\oplus
\Omega^{2, 0}(M)\oplus \Omega^{1, 1}(M)\subset\Omega^3(M)$. 

Let $\underline{\cO}_M$ denote the structure
sheaf of $M$. The exact sequence of
sheaves
$$
0\to \Z\to\underline{\cO}_M\stackrel{exp}{\to} \underline{\cO}^*_M
\to 0
$$
induces an exact sequence in cohomology
\begin{equation}\label{longexact}
\cdots\to H^2(M,\underline{\cO}_M)\to H^2(M,\underline{\cO}^*_M)
\stackrel{\delta}{\to}  H^3(M,\Z) \to H^3(M,\underline{\cO}_M)\to\cdots
\end{equation}
Since in general the cohomology groups $H^j(M,\underline{\cO}_M), \; j=2,
3,$  do not vanish, we conclude that holomorphic
bundle gerbes are certainly not classified by their Dixmier-Douady
invariant.

We have shown above that $\delta (H^2(M,\underline{\cO}^*_M))
\subset (H^{1,2}(M, \C)\oplus H^{2, 1}(M, \C))\cap 
H^3(M, \Z)$. We will now argue that in fact equality holds, that is 
 $\delta
(H^2(M,\underline{\cO}^*_M)) = (H^{1,2}(M, \C)\oplus H^{2, 1}(M, \C))\cap 
H^3(M, \Z)
$. (This result was known to I. M. Singer) To see this,
it suffices by (\ref{longexact}) to show that the image
of
$(H^{1,2}(M, \C)\oplus H^{2, 1}(M, \C))\cap 
H^3(M, \Z)$ in
$H^3(M,\underline{\cO}_M)$ is trivial. It suffices to show that the
image of
$H^{1,2}(M, \mathbb C)\oplus H^{2, 1}(M, \mathbb C)$ in
$H^3(M,\underline{\cO}_M)$ is trivial. Let $\pi_j : \Omega^{j}(M)
\to \Omega^{0,j}(M)$ denote the projection onto the subspace
of differential forms of type $(0,j)$. Then it is not hard to see that
the mapping
$$
H^3(M, \mathbb C) \to H^3(M,\underline{\cO}_M)
$$
is represented by a mapping of a $d$-closed differential 3-form $\psi$
onto the $\bar\partial$-closed differential form $\pi_3(\psi)$. Since any
class in $H^{i, j}(M, \mathbb C)$ is represented by a
$d$-closed differential form $\psi$ of type $(i,j)$,
it follows that
$\pi_3(\psi) = 0$ for all $d$-closed differential
form $\psi$ of type $(3,0)$,
$(1, 2)$ or $(2,1)$.In particular, the image of
$H^{1,2}(M, \mathbb C)\oplus H^{2, 1}(M, \mathbb C)$ in
$H^3(M,\underline{\cO}_M)$ is trivial as claimed.
We conclude that a bundle gerbe is stably isomorphic to a
holomorphic bundle gerbe if and only if its Dixmier-Douady
invariant lies in the subspace $(H^{1,2}(M, \C)\oplus H^{2, 1}(M, \C))\cap 
H^3(M, \Z)$ of $H^3(M,\Z)$.

\subsection{Holomorphic bundle gerbe modules.}

In this subsection, we will assume that $Y$ is a holomorphic 
principal $PGL(\mathcal H)$ bundle over $M$.
In the notation above, we consider holomorphic principal $GL_\K$
bundles $P$ over $Y$, and their associated Hilbert bundles $E$ over $Y$
in the standard representation. Such bundles have local holomorphic
trivializations and will be called holomorphic $GL_\K$-vector bundles
over $Y$. A holomorphic bundle gerbe module for $L$ is defined to
be a holomorphic $GL_{\K}$-vector bundle $E$ on $Y$ together with
an action of $L$ on $E$.
This is a holomorphic vector bundle
isomorphism
$\pi_1^*E\otimes L\to \pi_2^*E$ on $Y^{[2]}$
which is compatible with the product on $L$.
As in the previous cases there is
an extra condition regarding the action of $GL(\cH)$ on the
principal $GL_{\K}$ bundle associated to $E$.  As before we form
the holomorphic principal $GL(\cH)$ bundle $GL(E)$ on $Y$
with fibre at a point $y$ equal to the isomorphisms
$f\colon \cH\to E_p$.  Let $R$ denote the
principal $GL_{\K}$ reduction of $GL(E)$ determined by
the $GL_{\K}$ structure of $E$.  Then for $u\in GL(\cH)$ such
that $y_2 = y_1[u]$, the map $GL(E)_{y_1}\to GL(E)_{y_2}$
given by sending $f$ to $ufu^{-1}$ preserves $R$.
Analogous to the result in Section \ref{sec:four},
the space of holomorphic $GL_{\K}$-vector bundles on $Y$ forms a
semi-ring  under the operations of direct sum and tensor product.
It is easy to see that the operation of direct sum is compatible
with the action of the lifting holomorphic bundle gerbe $L\to
Y^{[2]}$  and so the set of holomorphic bundle gerbe modules for
$L$,
$\Mod_{GL_{\K}}^{hol}(L, M)$ has a natural structure as a
semi-group.   We denote the group associated to
$\Mod_{GL_{\K}}^{hol}(L, M)$  by the Grothendieck construction by
$\Mod_{GL_{\K}}^{hol}(L, M)$ as well.   We define the
(reduced) holomorphic bundle gerbe $K$-theory as
$$\tilde{K}_\varpi^0(M, Y) =
\Mod_{GL_{\K}}^{hol}(L, M),$$ where $[H]$ is the Dixmier-Douady
class of the holomorphic bundle gerbe $L\to Y^{[2]}$.
Moreover, we replace holomorphic $GL_{\K}$-vector
bundles by holomorphic $GL_{\mathrm{tr}}$-vector bundles
and recover the same holomorphic $K$-theory.  It can
be shown that
when $L$ is trivial, the holomorphic bundle gerbe $K$-theory
is isomorphic to the $K$-theory of holomorphic vector bundles as in
\cite{Hirz}.

By forgetting the holomorphic structure, we see that there
is a natural homomorphism $\tilde{K}_\varpi^0(M, Y)
\to \tilde{K}^0(M, Y)$ to bundle gerbe $K$-theory. By composing
with the  Chern character homomorphism $ch_{L}\colon \tilde{K}^0
(M, Y)\to H^{\ev}(M, Y)$ in bundle gerbe $K$-theory defined in
\cite{bcmms},  we obtain a Chern character homomorphism in
holomorphic bundle gerbe $K$-theory,
\begin{equation}\label{ch}
ch_{Y}\colon \tilde{K}_\varpi^0(M, Y)\to H^{\ev}(M, Y).
\end{equation}
It has the following  properties: 1) $ch_{Y}$ is natural
with respect  to pullbacks under holomorphic maps, 2) $ch_{Y}$
respects the $\tilde{K}_\varpi^0(M)$-module  structure of
$\tilde{K}_\varpi^0(M, Y)$ and 3)
$ch_{Y}$ reduces  to the ordinary Chern character on
$\tilde{K}_\varpi^0(M)$  when $Y$ is trivial, cf. \cite{Hirz}. 
Rationally, the image of the Chern character
(\ref{ch}) is far from being onto, as can be seen by choosing
hermitian connections compatible with the homomorphic structure in
the Chern-Weil description of the Chern character.  In the
particular case when
$Y$ is trivial, the image of the Chern character is contained in Dolbeault
cohomology classes of type $(p,p)$, and the precise image is
related to the Hodge conjecture. The Chern-Weil expression
for the Chern character in this context is again given by the
expression in Proposition 4.5.

\section{ Spinor bundle gerbe
modules}
\label{sec:spinC example}

In this section we give concrete examples of
bundle gerbe modules associated to a manifold
$M$ without a $Spin^{\C}$-structure and also to manifolds
without a $Spin$-structure. This construction
easily extends to the case when a general vector bundle
on $M$ does not either have a $Spin^{\C}$-structure or
a $Spin$-structure.

In the case when $M$ does not have a $Spin^{\C}$-structure,
this bundle gerbe module represents a class in twisted $K$-theory
of $M$, where the twisting is done by a 2-torsion
class in $H^3(M,\Z)$.  Suppose then that $M$ is an
$n$-dimensional oriented manifold without a $Spin^{\C}(n)$-structure.
It is a well known result of Whitney that orientable manifolds of
dimension less than or equal to four have $Spin^{\C}$-structures,
but there are many examples of higher dimensional
orientable manifolds that do not have any $Spin^{\C}$-structures.
One collection of examples of manifolds
that do not have any $Spin^{\C}$-structures are the Dold manifolds
$P(2m+1,2n)$, which is defined as the quotient $\left( S^{2m+1}
\times {\mathbb C} P^{2n}\right)/{\mathbb Z}_2$, where the action
of
${\mathbb Z}_2$ is given by $(x, z) \to (-x, \bar z)$.  Recall
that
$Spin^{\C}(n) = Spin(n)\times_{\Z_2} S^1$ and hence there is a
central extension
$S^1 \to Spin^{\C}(n)
\to SO(n)$. Let $SO(M)$ denote the oriented frame bundle of $M$.
Then associated to $SO(M)$ is the lifting bundle gerbe
arising from the central extension of $Spin^{\C}(n)$.
More precisely over the fibre product $SO(M)^{[2]} =
SO(M)\times_{\pi}SO(M)$ (here $\pi\colon SO(M)\to M$
denotes the projection) we have the canonical map
$SO(M)^{[2]}\to SO(n)$.  We can pullback the principal
$S^1$-bundle $Spin^{\C}(n)\to SO(n)$ to $SO(M)^{[2]}$ via
this map.  The resulting bundle $L\to SO(M)^{[2]}$ is a
bundle gerbe. It is natural to call this a $Spin^{\mathbb C}$
bundle gerbe.  The
Dixmier-Douady class of this bundle gerbe in
$H^3(M,\Z)$  coincides with the third integral Stieffel-Whitney class
$W_3(TM)$, which measures precisely the obstruction to $M$ being
$Spin^{\C}(n)$. Recall that
$W_3(TM) = \beta w_2(TM)$, the Bockstein $\beta$ applied
to the second Steifel-Whitney class $w_2(TM)$.  As a
consequence $W_3(TM)$ is a $2$-torsion class.

We can pullback the $SO(n)$-bundle $SO(M)\stackrel{\pi}{\to} M$
to $SO(M)$ via $\pi$ to get an $SO(n)$ bundle $\pi^*SO(M)\to SO(M)$.
Since $W_3(\pi^*SO(M)) = 0$,  we can construct a lift
$\widehat{\pi^*SO(M)}$ of the  structure group of $\pi^*SO(M)$ to
$Spin^{\C}(n)$; i.e. there  is a bundle map $\widehat{\pi^*SO(M)}\to
\pi^*SO(M)$  covering the homomorphism $p\colon Spin^{\C}(n)\to SO(n)$.
It is easy to see what this lift is.  Note that
$\pi^*SO(M)$ identifies canonically with $SO(M)^{[2]}$.
Therefore we can regard the line bundle $L\to SO(M)^{[2]}$
as sitting over $\pi^*SO(M)$.  It is easy to see that a lift
$\widehat{\pi^*SO(M)}$ is given by the total space of the line bundle
over $\pi^*SO(M)$.

We can form the bundle of spinors $\cS=\cS(\widehat{\pi^*SO(M)})
\to SO(M)$
associated to
$\widehat{\pi^*SO(M)}$.  Recall that we do this by taking an
irreducible representation $V$ of $Spin^{\C}(n)$ and
forming the associated bundle $\cS =
\widehat{\pi^*SO(M)}\times_{Spin^{\C}(n)} V$ on $SO(M)$.
It is straightforward to show that $\cS$ is a module
for the bundle gerbe $L\to SO(M)^{[2]}$.  Recall that in the
case when the dimension of $M$ is even, there are two half spin
representations and in the odd dimensional case, there is a unique
spin representation.  It is natural to call this a spinor
bundle gerbe module.

This discussion can be extended to cover the case when we
have a real vector bundle $E\to M$.  We can then associate a
lifting bundle gerbe to the oriented frame bundle $SO(E)$
which measures the obstruction to $SO(E)$ having a
lift of the structure group to $Spin^{\C}$.
On $SO(E)$ we can form the pullback bundle $\pi^*SO(E)$.
This has a lift $\widehat{\pi^*SO(E)}$ to a $Spin^{\C}(n)$
bundle on $SO(E)$.  Associated to $\widehat{\pi^*SO(E)}$ we
can construct the bundle of spinors $\cS(\widehat{\pi^*SO(E)}) \to SO(E)$:
again  this is a module for the lifting bundle gerbe
$L\to SO(E)^{[2]}$.

The possible spinor bundle gerbe modules for
the $Spin^{\mathbb C}$ bundle gerbe $L\to P^{[2]}$
are parametrised by
$H^2(M,\Z)$.  We see this as follows.  
Let $\cL\to M$ be a line bundle on
$M$.  Then given a spinor
bundle gerbe module $\cS$ we can extend the action of $L$ on $\cS$
to an action on $\cS\otimes \pi^*\cL$ by acting trivially with $L$ on
$\pi^*\cL$.  $\cS\otimes \pi^*\cL$ is a spinor bundle gerbe
module since tensoring with $\pi^*\cL$ preserves rank and
therefore takes bundles of irreducible $Spin^{\C}$ modules to bundles of
irreducible $Spin^{\C}$ modules.  Thus we have an action of 
the category of line bundles on $M$  
on the category of all spinor bundle gerbe modules.  The following
argument showing that this is a transitive action is due to M.~Murray.
Suppose we have two spinor bundle
gerbe modules $\cS_1$ and $\cS_2$ on $SO(M)$.  The frame
bundles of $\cS_1$ and $\cS_2$ provide lifts $\widehat{\pi^*SO(M)}_1$
and $\widehat{\pi^*SO(M)}_2$ respectively of the structure
group of $\pi^*SO(M)$ to $Spin^{\C}(n)$.  It is straightforward
to see that there is a principal $U(1)$ bundle $P$ on
$SO(M)$ such that $\widehat{\pi^*SO(M)}_2 = \widehat{\pi^*SO(M)}_1
\otimes P$.  Let $\hat{\cL}$ denote the complex line on
$SO(M)$ associated to $P$.  Then
we have an isomorphism $\cS_2 =
\cS_1 \otimes \hat{\cL}$.  We claim that the line bundle
$\hat{\cL}$ on $SO(M)$ descends to a line bundle $\cL$ on
$M$.  To see this, think of the fibre $\hat{\cL}_y$ of
$\hat{\cL}$ at $y\in SO(M)$ as consisting of isomorphisms
$(\cS_1)_y \to (\cS_2)_y$.  Similarly
we can think of a point $l$ of the fibre
$L_{(y,y')}$ of $L$ at $(y,y')\in SO(M)^{[2]}$ as
an isomorphism $l\colon (\cS_1)_{y'} \to (\cS_1)_y$
or as an isomorphism $l\colon (\cS_2)_{y'}\to (\cS_2)_y$.
We can therefore use $l$ to construct an isomorphism
$\hat{\cL}_{y'} \to \hat{\cL}_y$.  This isomorphism
is easily seen to be independent of the choice of
$l\in L_{(y,y')}$.  It therefore follows that the
line bundle $\hat{\cL}$ descends to a line bundle
$\cL$ on $M$.

The construction above can be generalized in a straightforward
manner to yield the following proposition, which constructs
bundle gerbe modules in the case when the Dixmier-Douady invariant
is a torsion class.

\begin{proposition}
(1) The following data can be used to construct bundle
gerbe modules $E$ on a manifold $M$.
\begin{itemize}
\item A principal $G$-bundle $P$ on $M$, where $G$ is a finite
dimensional Lie group;
\item A central extension
$$
U(1)\to \widehat G \to G;
$$
\item A finite dimensional representation
$$
\rho: \widehat G \to GL(\cH)
$$
such that the restriction of $\rho$ to the central $U(1)$ subgroup 

is the identity.
\end{itemize}
The bundle gerbe modules $E$ determine elements in $K^0(M,
P)$.


(2) The following data can be used to
construct holomorphic bundle gerbe modules $E$ on a complex
manifold $M$.
\begin{itemize}
\item A holomorphic principal $G$-bundle $P$ on $M$ where $G$
is a finite dimensional complex Lie group;;
\item A central extension
$$
{\mathbb C}^*\to \widehat G \to G;
$$
\item A finite dimensional representation
$$
\rho: \widehat G \to GL(\cH).
$$
such that the restriction of $\rho$ to the central ${\mathbb C}^*$ 
subgroup 
is the identity.
\end{itemize}
The holomorphic bundle gerbe modules $E$ determine elements in
$K^0_\varpi(M,  P)$.


(3) Suppose that $K$ is a compact, connected and
simply connected simple Lie group.  Then the following
data can be used to construct $K$ equivariant bundle
gerbe modules $E$ on a manifold $M$.
\begin{itemize}
\item A $K$ equivariant principal $G$-bundle $P$ on $M$,
where $G$ is a finite dimensional Lie group;
\item A central extension
$$
U(1)\to \widehat G \to G;
$$
\item A finite dimensional representation
$$
\rho: \widehat G \to GL(\cH).
$$
such that the restriction of $\rho$ to the central $U(1)$ subgroup 

is the identity.
\end{itemize}
The $K$ equivariant bundle gerbe modules $E$ determine elements in
$K^0_K(M, P)$.

\end{proposition}

In part $(3)$ of the Proposition, the only difficulty
lies in showing that the class in $H^3_K(M;\Z)$
associated to the $K$-equivariant principal $G$-bundle
$P\to M$ vanishes when lifted to $H^3_K(P;\Z)$.  To
see this it suffices to consider the universal case,
when $P$ is $EG$.  Recall that the $K$-equivariant cohomology
of $EG$ is defined to be the cohomology of the space
$EG\times_K EK$.  Note that $K$ acts freely on $EG\times
EK$ and hence $EG\times_K EK$ has the homotopy type of
$BK$.  The hypotheses on $K$, namely $1=\pi_0(K)=\pi_1(K) = \pi_2(K)$
imply that $1=\pi_0(BK)=\pi_1(BK)=\pi_2(BK) = \pi_3(BK)$, using the
fibration $K\to EK\to BK$ and the long exact sequence in homotopy.
By the Hurewicz theorem, it follows that $0 =
H_1(BK;\Z)=H_2(BK;\Z) = H_3(BK;\Z)$. By the universal coefficient
theorem, it follows that $H^3(BK;\Z) = 0$, that is, the degree  three
$K$-equivariant cohomology of
$EG$ vanishes.
It  follows that we can then choose a $K$-equivariant
lift $\widehat{\pi^*P}$ of $\pi^*P$ to a $\hat{G}$ bundle.
We can then form the associated vector bundle $E = \widehat{
\pi^*P}\times_{\rho}\cH$.  It is clear that $E$ is a
$K$-equivariant bundle gerbe module.

Proposition 8.1 can be viewed as the analogue in the
twisted case of the
associated bundle construction. It can be formalised as
follows. Let $G, \widehat G$ be a compact Lie group 
and a $U(1)$ 
central extension respectively. 
Let $R(\widehat G)$ denote the 
representation ring of 
 $\widehat G$, which we recall is defined
as the free Abelian group generated by the irreducible
complex representations of $\widehat G$. Let $R_0(\widehat G)$
denote 
the subgroup of $R(\widehat G)$ defined as those 
representations 
$\rho$ of $\widehat G$ 
such that the restriction of $\rho$ to the 
central $U(1)$ subgroup 
is the identity.

The augmentation homomorphism
$\varepsilon : R_0(\widehat G) \to \Z$ assigns to each representation
in $ R_0(\widehat G)$ its dimension, and the augmentation subgroup
 $I_0(\widehat G)$ is
the kernel of $\varepsilon$. 
Given a principal $G$ bundle over $M$, the construction in Proposition
8.1 part (1) yields a  homomorphism
$$
\alpha_P : R_0(\widehat G) \to K^0(M,  P).
$$
If $M$ is a point, then the homomorphism reduces to the augmentation
homomorphism $\varepsilon$.
If $f: N\to M$ is a smooth map, then it is not hard to see that 
the following diagram commutes,
$$
\CD
K^0(M, P) \;\; \overset{f^!}{\longrightarrow}\;\;K^0(N,
f^*P)\\
  \overset{\alpha_P}{\nwarrow} \qquad
\overset{\alpha_{f^*P}}\nearrow\\ R_0(\widehat
G).
\endCD
$$
Similarly,  the hypotheses of Proposition 8.1 part (3) yields
the homomorphism
$$
\alpha_P : R_0(\widehat G) \to K^0_K(M,  P).
$$

We now consider the case when the manifold $M$ does
not have a $Spin$-structure.
The discussion above also makes sense if we replace
$Spin^{\C}(n)$ by $Spin(n)$ and consider
the central extension $\Z_2 \to
Spin(n)\to SO(n)$, so we will avoid repetition.  Given a
principal
$SO(n)$  bundle $P$ on $M$ (in particular the oriented bundle of
frames  on $M$) we can
consider the lifting bundle  gerbe associated to this central
extension of
$SO(n)$  by $\Z_2$.  This time we will have a principal
$\Z_2$ bundle $L\to P^{[2]}$ (or equivalently a real
line bundle over $P^{[2]}$). It is natural to
call this a $Spin$ bundle gerbe.  The
`real version' of the Dixmier-Douady invariant of the $Spin$ bundle
gerbe coincides with the second Stieffel-Whitney class of $P$ in
$H^2(M, {\mathbb Z}_2)$. We remark that the real version of
Dixmier-Douady theory involves the obvious modifications to
standard Dixmier-Douady theory, and is in the
literature (cf. \cite{Ros}). The application of
the real version of Dixmier-Douady theory to the real version
of bundle gerbe theory is what is used here, the
details of which are obvious modifications of the
standard theory of bundle gerbes.
As above, the pullback
$\pi^*P$ of $P$ to $P$ has a lifting to a $Spin(n)$
bundle $\widehat{\pi^*P}\to P$.  We consider the associated
bundle of spinors by taking an irreducible representation $V$
of $Spin(n)$ and forming the associated vector bundle
$\cS = \widehat{\pi^*P}\times_{Spin(n)} V$ on $P$.  $\cS$ is a
bundle gerbe module for $L$, called a spinor bundle gerbe module
as before.
One can show that the possible  spinor bundle gerbe modules for
the $Spin$ bundle gerbe $L\to P^{[2]}$ are parametrised by
$H^1(M,\Z_2)$, i.e. the real line bundles  on $M$, by following
closely the proof given above in the
$Spin^{\mathbb C}$ case.


\end{document}